\documentclass[a4paper,11pt]{article}
\pdfoutput=1 % if your are submitting a pdflatex (i.e. if you have
\usepackage{jcappub} % for details on the use of the package, please
\usepackage{soul}

\usepackage[T1]{fontenc} % if needed
\usepackage{natbib}
\usepackage{subcaption}

\title{Limits on self-interacting neutrinos from the BAO and CMB phase shift}

\author[a,1]{Abbé M. Whitford,\note{Corresponding author.}}
\author[a]{Cullan Howlett,}
\author[a]{Tamara M.\ Davis,}
\author[b]{David Camarena,}
\author[b]{Francis-Yan Cyr-Racine}

\affiliation[a]{The University of Queensland,\\St Lucia QLD, Australia}
\affiliation[b]{Department of Physics and Astronomy, University of New Mexico,\\210 Yale Blvd SE, Albuquerque NM, United States}

\emailAdd{abbe.whitford@gmail.com}
\emailAdd{c.howlett@uq.edu.au}
\emailAdd{tamarad@physics.uq.edu.au}
\emailAdd{dcamarena93@unm.edu}
\emailAdd{fycr@unm.edu}

\abstract{Neutrinos with Standard Model interactions free-stream in the early Universe, leaving a distinct phase shift in the pattern of baryon acoustic oscillations (BAO). When isolated, this phase shift allows one to robustly infer the presence of the cosmic neutrino background in BAO and cosmic microwave background (CMB) data independently of other cosmological parameters. While in the context of the Standard Model, this phase shift follows a known scale-dependent relation, new physics in the cosmic neutrino background could alter the overall shape of this feature. In this paper, we discuss how changes in the neutrino phase shift could be used to constrain self-interactions among neutrinos. We produce simple models for this phase-shift assuming universal self-interactions, and use these in order to understand what constraining power is available for the strength of such interactions in BAO and CMB data. We find that, although challenging, it may be possible to use a detection of the phase to put a more robust limit on the strength of the self-interaction, $G_{\mathrm{eff}}$, which at present suffers from bimodality in cosmological constraints. Our forecast analysis reveals that BAO data alone will not provide the precision needed to tightly constrain self-interactions; however, the combined analysis of the phase shift signature in both CMB and BAO can potentially provide a way to detect the impact of new neutrino interactions. Our results could be extended upon for models with non-universal interactions.}

\begin{document}
\maketitle
\flushbottom

\section{Introduction }
\label{sec:int}

In the early Universe, neutrinos quickly decouple from the hot plasma of baryons and photons that allowed sound waves to propagate prior to recombination. These free-streaming neutrinos are referred to as the cosmic neutrino background (C$\nu$B). The presence of the C$\nu$B impacts the cosmic microwave background (CMB) photons and also the growth of structure we see in the Universe today \citep[see detailed reviews in][]{lesgourgues2006massive, lesgourgues2012neutrino, lesgourgues2014neutrino}. While the C$\nu$B itself has not been directly detected due to the nature of neutrinos as very weakly-interacting particles, the impacts of the C$\nu$B on cosmological observables have been observed in the CMB \citep{follin2015first, aghanim2020planck, montefalcone2025free}, the matter power spectrum \citep{elgaroy2002new}, the Lyman-$\alpha$ forest \citep{yeche2017constraints}, measurements of baryon acoustic oscillations \citep[BAOs,][]{baumann2018searching, baumann2019first, adame2024desi, whitford2024constraining, elbers2025constraints, karim2025desi} and combinations of cosmological data \citep{tegmark2004cosmological}, including terrestrial experiments \citep{stocker2021strengthening}. Big Bang Nucleosynthesis (BBN) is also able to help constrain neutrino properties \citep{mangano2011robust}. These observations constrain the sum of the masses of the different neutrino species $\sum m_{\nu}$ and the effective number of neutrino species, $N_{\mathrm{eff}}$. The latter parameterises the number of neutrino species and their contribution to the radiation density $\Omega_{\mathrm{r}}$ via 
\begin{align}
    \Omega_{\mathrm{r}} & = \Omega_{\nu} + \Omega_{\gamma} \\ \nonumber
    & = \Omega_{\gamma}\left( 1 + N_{\mathrm{eff}} \frac{7}{8} \left(\frac{4}{11}\right)^{4/3} \right).
\end{align}
$N_{\mathrm{eff}}$ is expected to be 3.044 in the Standard Model.\footnote{While there are expected to be just three neutrino species or three mass eigenstates, $N_{\mathrm{eff}}$ increases to 3.044 to absorb some of the additional energy that is injected into the neutrino distribution during electron-positron annihilation in the early Universe~\citep{deSalas:2016ztq,Bennett:2020zkv, akita2020precision}.} As such, altering $N_{\mathrm{eff}}$ alters the radiation density, altering the expansion rate in the early Universe. Larger $N_{\mathrm{eff}}$ decreases the distance of the sound horizon at recombination, captured by the parameter $\theta_s$ in CMB data, and decreases the sound horizon distance of the baryon drag epoch $r_s$, measured in BAO data. These changes to the expansion rate also impact Silk damping and the growth of structure on scales that enter the horizon prior to matter-radiation equality, as the growth of structure is suppressed in the radiation-domination era.
%the amount of time that photons can execute a random walk that washes out temperature perturbations on small scales during the baryon acoustic oscillations, via a process known as diffusion damping. Changes to the radiation density also impact on structure growth on small scales which are suppressed during the radiation-domination epoch in the early Universe. 
This is reflected in the matter power spectrum and the CMB power spectra, allowing for $N_{\mathrm{eff}}$ constraints. Free-streaming neutrinos in the early Universe also induce a distinct imprint on the evolution of the photons and matter perturbations. This imprint, notable as a phase shift in both the CMB and matter power spectrum,  is a unique signature of the C$\nu$B and can be used to robustly detect the presence of free-streaming relics \citep{follin2015first, baumann2019first, wallisch2019cosmological, montefalcone2025free} or the lack thereof, if there are instead fluid-like relics that do not free-stream \citep{montefalcone2025free}.

\subsection{Searching for neutrino physics beyond the Standard Model}

Despite these results, neutrinos themselves are still one of the most poorly understood particles in the Standard Model as, for instance, it is not entirely clear how they acquire their mass.  While terrestrial oscillation experiments are able to constrain the squared mass differences between different neutrino species \citep{fukuda1998evidence, hirata1989observation},\footnote{These experiments measure mass splittings $\Delta m_{ij}^2$ between different neutrino eigenstates labelled by $i,j$, which puts a lower bound on $\sum m_{\nu}$. } and weak constraints on neutrino masses can be obtained via beta decay or neutrinoless double beta decay \citep{tanabashi2018review, particle2020review, vitagliano2020grand}, the tightest upper bounds on neutrino mass to date come from cosmological data (via constraints on $\sum m_{\nu}$ \citep{elbers2025constraints, adame2024desi}). However, some of these measurements have given extremely puzzling results, with a preference for a zero or negative $\sum m_{\nu}$ found in the data \citep{craig2024no, green2024cosmological, adame2024desi}, although such results are somewhat relaxed by the analysis presented in ref.~\cite{elbers2025constraints}.

Beyond neutrino masses, cosmological data can also provide some insight into new physics coupling to neutrinos. Indeed, the early Universe, with its high density and pressure, may allow for the production of yet-unknown weakly coupled light particles \cite{baumann2018searching}, allowing cosmology to probe interactions that cannot be accessed by terrestrial experiments. The presence of such additional light species (including sterile neutrinos) would change $N_{\mathrm{eff}}$ in a way that would be potentially detectable in cosmological data. It may also be possible for cosmological data to search for hints of non-standard interactions in neutrinos. 

The parameter $N_{\mathrm{eff}}$ captures the effects of light relics which contribute to the radiation density $\Omega_{\mathrm{r}}$, including those that free-stream, but also the effects that may be caused by species that are fluid-like (which do not free-stream). As such, in ref.~\cite{montefalcone2025free}, an approach is taken to isolate and separately constrain impacts of free-streaming and fluid-like relics. Similarly, the impacts of dark radiation (which may involved fluid-like components, free-streaming components or radiation with varying decoupling or recoupling mechanisms) are studied in \cite{brinckmann2023confronting, saravanan2025abundance}. In other works, there have been searches for signatures of non-standard neutrino interactions in the matter power spectrum and CMB power spectra. In particular, refs.~\cite{kreisch2020neutrino, camarena2023confronting, kreisch2024atacama, poudou2025self, he2024self, park2019lambda, oldengott2017interacting, venzor2022massive} have studied the impacts of self-interacting neutrinos on the CMB and matter power spectra, and have found potential hints of strong neutrino self-interactions in the early Universe. For example, ref.~\cite{poudou2025self} find that DESI BAO \citep{adame2024desi} data combined with CMB data from the new \textit{Planck} analysis \citep{tristram2024cosmological, rosenberg2022cmb} prefers neutrinos with strong self-interactions over $\Lambda$CDM, with the difference between the minimum $\chi^2$ in each case as $-4.3$ (this comparison involves 8 model parameters + $G_{\mathrm{eff}}$ in the strong self-interaction model). Other analyses have found that Standard Model neutrinos are still preferred by the data despite such hints \citep{lancaster2017tale, brinckmann2021self, choudhury2021updated, camarena2025strong, he2025fresh}.

\subsection{Studying the phase shift of self-interacting neutrinos: motivation}

Searching for signatures of self-interacting neutrinos in cosmological data was initially motivated by the Hubble and $\sigma_8$ tensions \citep{kreisch2020neutrino, poudou2025self, berryman2023neutrino, berbig2020hubble}, since allowing for neutrinos that self-interact (or otherwise have delayed free-streaming, a consequence of neutrino self-interactions) alters the sound horizon scale measured from the CMB or BAO data. However, the search for new neutrino interactions are also motivated by the need to explain anomalous results in terrestrial neutrino experiments \citep{crivellin2024anomalies} and even perhaps to explain the apparent preference for negative neutrino mass in cosmological analyses \citep{craig2024no, das2025impostor} or understand constraints in the studies in which additional freedom is allowed in neutrino properties \citep{pal2025exploring}. Self-interactions can also allow for greater flexibility in fits to inflationary models \citep{choudhury2022massive, barenboim2019constraints}. The self-interactions in these works generally assume a model in which all neutrinos can share a common parameter to describe the strength of self-interactions, $G_{\mathrm{eff}} = |g_{\nu}|^2/m^2_{\phi}$, for a four-fermion contact interaction. The universal coupling is given by $g_{\nu}$ and the mediator mass is given by $m_{\phi}$. While several terrestrial experiments \citep{lyu2021self, blinov2019constraining} put strong constraints on this kind of interaction, modelling this interaction in cosmological data can capture the general impacts that neutrinos with more complicated interactions may have, thus it is useful to study. For example, ref.~\cite{das2025impostor} study a scenario in which neutrinos convert into dark radiation after BBN but before recombination. This leaves a signal similar to strongly self-interacting neutrinos in cosmological data, without violating constraints from terrestrial experiments. 

However, in many cases, the study of neutrino self-interactions through the $G_{\mathrm{eff}}$ parameterization leads to bimodal constraints. In various works \citep[see for example][]{camarena2023confronting, camarena2025strong, kreisch2020neutrino, kreisch2024atacama, berryman2023neutrino, lancaster2017tale, oldengott2017interacting, park2019lambda} one sees two peaks in the posterior for $log_{10}{(G_{\mathrm{eff}})}$; one peak shows a preference for very strong self-interactions with $log_{10}{(G_{\mathrm{eff}})} ~ -1.5$. The second peak prefers sufficiently weaker interactions that are not as strongly constrained and the simultaneous fits to other cosmological parameters are somewhat consistent with $\Lambda$CDM. The $\Lambda$CDM model and the weaker neutrino self-interaction model are not so clearly distinguished. Due to the nature of the impact that neutrino self-interactions have on the broadband-shape of the matter power spectrum, it is difficult to distinguish whether the peak associated with stronger interactions is simply a result of degeneracies with the parameters $n_s$ and $A_s$ \citep{berryman2023neutrino} or truly a hint for strong-self interactions. However, the impact that neutrinos have on the phase of the BAO and CMB peaks is a more distinct feature that could be studied that is induced by free-streaming relics. 

In this paper, we continue to explore how cosmological data can be used to search for signatures of neutrinos with non-standard interactions, going beyond a simple measurement of $N_{\mathrm{eff}}$. In particular, we explore the (relative absence of) phase shift that neutrinos with self-interactions leave in the CMB and BAO oscillations of the matter power spectrum. We study this in an effort to understand what information is available to constrain non-standard interactions in neutrino species, which is expected to alter the phase shift signal compared to that seen for Standard Model neutrinos. In particular, this may be useful given the bimodal constraints on strongly self-interacting neutrinos seen in the literature. As mentioned, the phase shift in the CMB and BAO is uniquely created by free-streaming relics, and thus is not degenerate with $A_s$ or $n_s$ alone. In section~\ref{sec::selfinteractionsintro}, we further discuss the neutrino self-interaction model, the cosmological impacts of these interactions and show the phase shift signal in the CMB and BAO, motivating the study of this signal. In section~\ref{sec::standardmodelphase}, we revisit the existing templates for the phase shift signal of Standard Model neutrinos that have been constructed in CMB and BAO data. We build upon these templates in section~\ref{sec::selfinteractiontemplate} to construct a template for the phase shift that captures the impact of neutrinos with self-interactions. In sections~\ref{sec::constraintsBAO}, \ref{sec:constraintsCMB}, we explore the ability to constrain $G_{\mathrm{eff}}$ in the BAO and CMB, and conclude in section~\ref{sec:conclusion}. Throughout this work we generally keep $\Omega_{\mathrm{b}} = 0.048$, $h = 0.6763$, $\tau = 0.066$, $n_s = 0.9667$, $\ln{A_s10^{10}} = 3.064$ unless stated otherwise. To extract the phase shift signal, it is necessary to vary $N_{\mathrm{eff}}$ away from a fiducial value of $3.044$, and $\Omega_{\mathrm{m}}$ is varied about the fiducial value of $0.31$ to keep the matter-radiation equality scale fixed. For the CMB, it is necessary to alter $Y_p$ about a fiducial value of $0.245$ to keep the Silk-damping scale fixed. 

% \section{Exploring the phase shift with neutrino self-interactions}

\section{Cosmological impacts of neutrino self-interactions}\label{sec::selfinteractionsintro}

The impact of flavor-universal neutrino self-interactions---parameterized by $\log_{10}{(G_{\mathrm{eff}})}$, with $G_{\mathrm{eff}}$ in units of MeV$^{-2}$ ---on the matter power spectrum arises from their impact on neutrino free-streaming. In the Standard Model, neutrinos free-stream from very early times, approximately one second after the Big Bang.\footnote{For \emph{massive} neutrinos, they transition from relativistic particles to non-relativistic paticles as their thermal velocities decay. As such, the free-streaming has the impact of suppressing small-scale power by carrying away energy, while clustering on scales larger than their free-streaming length. However, this is an impact that occurs much later and is essentially unaffected by their decoupling from other particles (or even delayed free-streaming is they remain coupled for longer) unless the epoch of free-streaming is significantly delayed. This affect is captured by changes to $\sum m_{\nu}$.} The phase shift that is caused by free-streaming for Standard Model neutrinos is discussed in more detail in the following section, section~\ref{sec::standardmodelphase}. 

Strong self-interactions delay this free-streaming, altering the gravitational potential on different length scales. Consequently, strong self-interactions also alter the phase shift signal of free-streaming neutrinos. We describe how this is so more in the next paragraphs and in section~\ref{sec::selfinteractiontemplate} where we show the template for the phase shift. To first give context to this, in these self-interaction scenarios, the matter power spectrum shows enhanced power on scales that enter the horizon when self-interactions freeze out and the onset of free streaming begins. Conversely, a suppression on smaller scales appears on scales that enter the horizon while neutrinos are still tightly coupled. In this way, a scale-dependent feature is imprinted, producing effects that are somewhat degenerate with $A_s$ (amplitude) and $n_s$ (tilt).

\begin{figure}
    \centering
    \includegraphics[width=\linewidth]{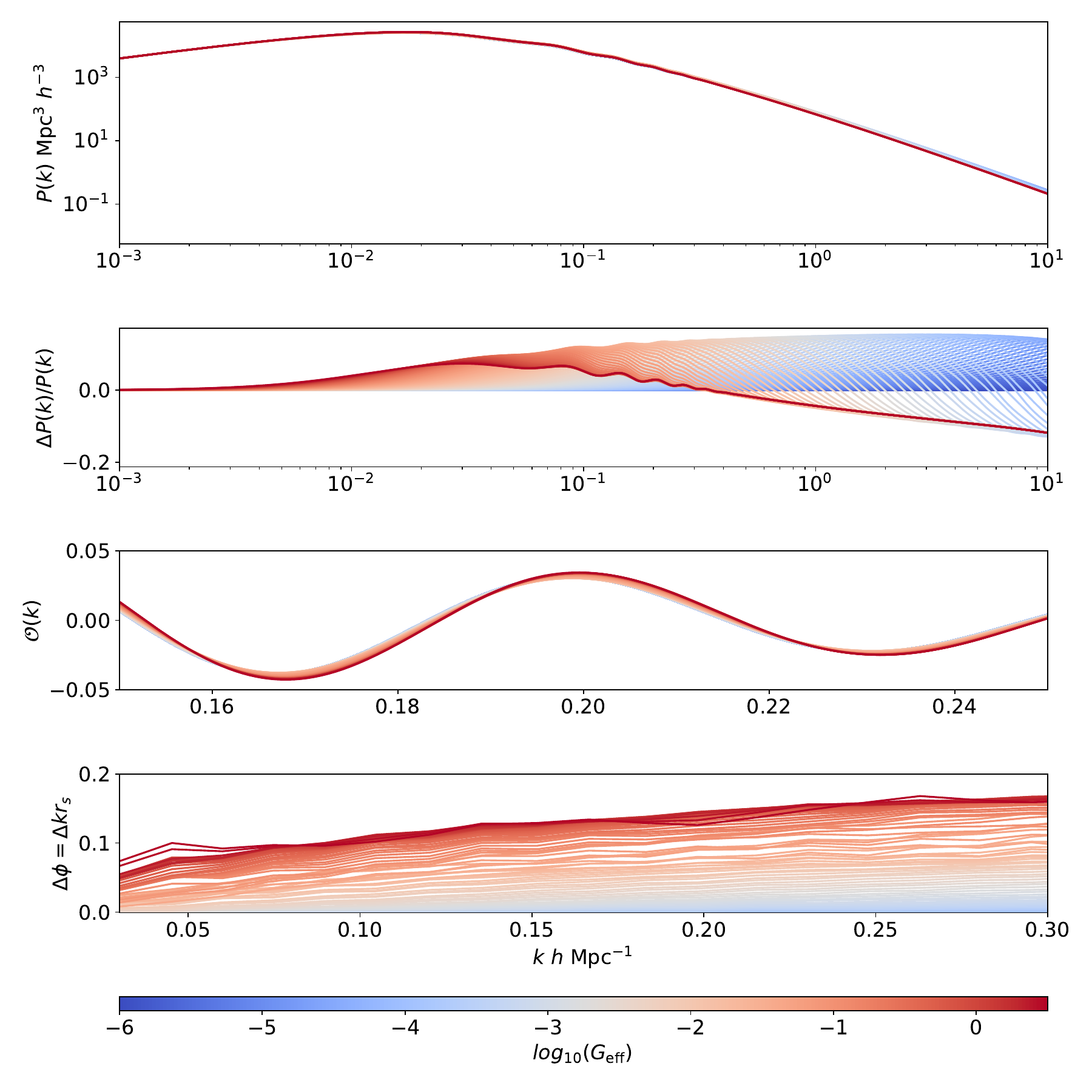}
    \caption{The impact of $G_{\mathrm{eff}}$ (the interaction strength of neutrinos) on the matter power spectrum. This is directly shown in the first panel, with the relative change to a power spectrum with $\log_{10}{(G_{\mathrm{eff}})} = -6$ (effectively indistinguishable to $\Lambda$CDM) shown in the next panel. In the third panel, we show the BAO oscillations computed by dividing out the smoothed broadband power spectrum. We also alter $A_s$ and $n_s$ before implementing the smoothing procedure, which is explained in further detail in the main text. The shift in the phase $\Delta \phi = \Delta k r_s$ as a function of $k$ is computed as the shifts in the peaks and troughs of the power spectrum oscillations relative to a power spectrum with $\log_{10}{(G_{\mathrm{eff}})} = -6$; this is shown in the fourth panel. It is possible to also see this signal as a function of $\ell$ in the CMB power spectrum, by plotting $\Delta \ell$ (or $\theta_s \Delta \ell$); see Figure~\ref{fig:phaseshiftdemo_CMB}. }
    \label{fig:phaseshiftdemo_BAO}
\end{figure}

\begin{figure}
    \centering
    \includegraphics[width=1\linewidth]{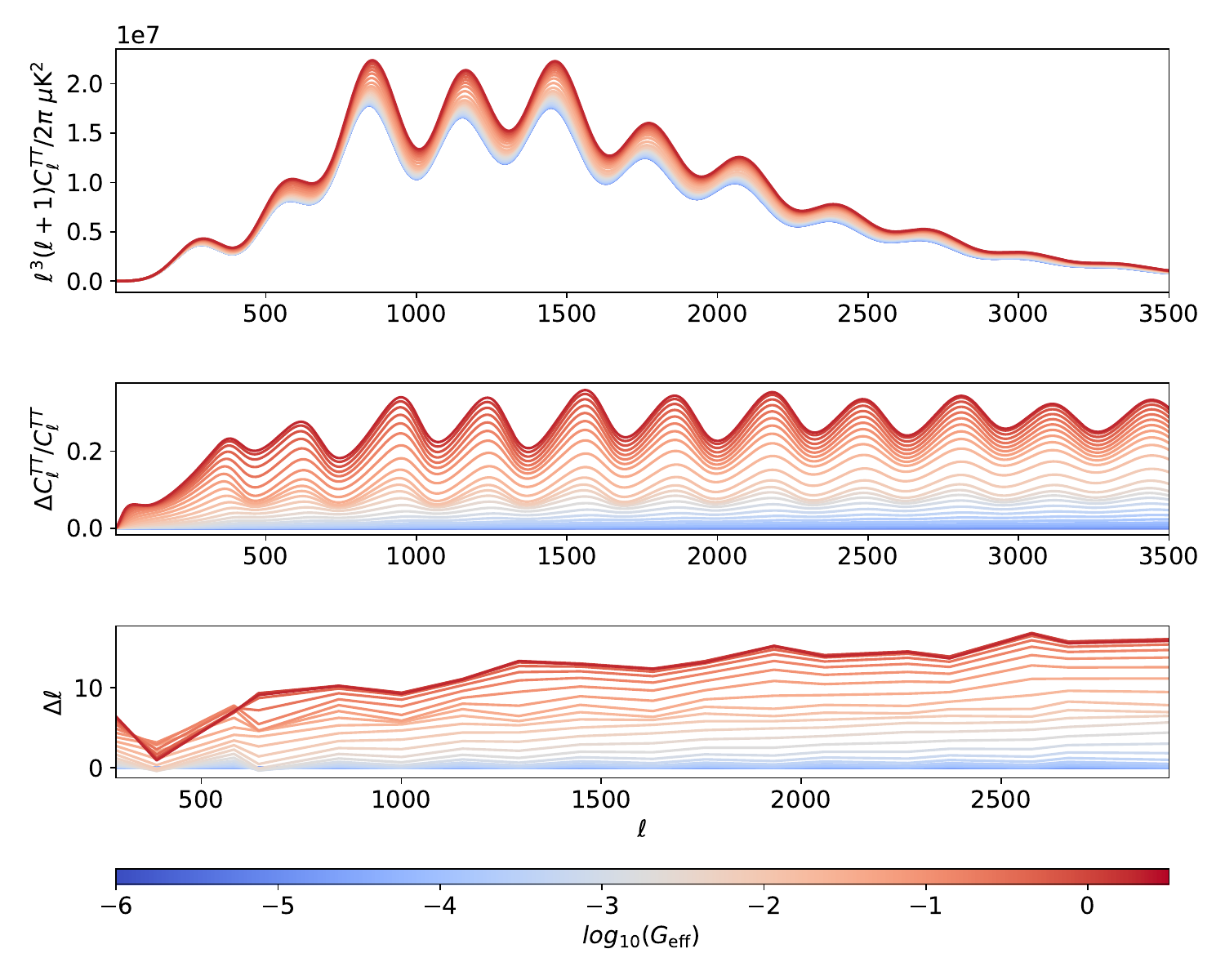}
    \caption{The effect of neutrino self-interactions and thus delayed free-streaming on the CMB temperature-temperature (TT) power spectrum; this is directly shown by plotting power spectra with various $G_{\mathrm{eff}}$ in the top panel, with the relative difference shown in the second panel. The final panel shows the shift in the peaks and troughs of the CMB oscillations, $\Delta \ell$, which is the phase shift caused by neutrino self-interactions.}
    \label{fig:phaseshiftdemo_CMB}
\end{figure}

The additional consequence of modifying the epoch of neutrino free-streaming is that this also modifies the characteristic phase shift observed in power spectrum oscillations. For Standard Model neutrinos, the phase shift approaches zero on the largest scales but has an almost constant amplitude on smaller scales; the amplitude is approximately independent of cosmology except for $N_{\mathrm{eff}}$; for larger $N_{\mathrm{eff}}$ there is a larger net amplitude for the phase shift. When free streaming is delayed due to self-interactions, neutrinos shift the phase of photon–baryon plasma waves only for modes that cross the horizon after these interactions cease to a first approximation, leading to a scale-dependent change to the oscillation phase. The detailed form of this effect has been less studied in the context of self-interactions and is the focus of this paper. Figures~\ref{fig:phaseshiftdemo_BAO} and \ref{fig:phaseshiftdemo_CMB} illustrate how the matter and CMB temperature-temperature power spectra and their corresponding phase shift effects change when self-interactions among neutrinos are considered.

While the impact of delayed neutrino free-streaming in the context of universal neutrino self-interactions on the phase shift has not been studied extensively, some similar scenarios have been explored in some detail. Ref.~\cite{choi2018probing} shows the results of analytical and numerical calculations of a phase shift in the CMB power spectra (TT, EE) when an additional species contributes $\Delta N_{\mathrm{eff}}$. In that work, the additional species are given delayed free-streaming compared to Standard Model neutrinos. They show that the phase and amplitude of CMB power spectra (TT, EE) are affected, with a dependence on $\Delta N_{\mathrm{eff}}$ and the epoch at which free-streaming occurs. The decoupling species induces a scale-dependent change in the phase shift and leads to a drop in the phase signal amplitude (relative to the Standard Model neutrino phase shift) on small scales. These scales correspond to those smaller than the horizon when the species decouple. More recently, \cite{montefalcone2025directly} has studied the phase shift for neutrinos that have self-interactions or dark-matter couplings, causing neutrinos to behave as a fluid prior to decoupling and propagating faster than the sound speed. They find in the case of gradual decoupling of neutrinos from self-interactions or dark-matter couplings over a finite length of time, the phase shift has a roughly constant amplitude dependence, with a small scale-dependence that they choose to not include in their modelling or analysis since the effect is small.  In our own analysis we include a subtle scale dependence in the phase shift measurements, which may be detectable with highly sensitive future experiments. Additionally, \cite{saravanan2025abundance} studies in some detail the impact of altering the radiation density and the fraction of it which free-streams or is instead fluid-like; we expect that the impact on the phase shift of increasing the fluid-like radiation density should be similar to increasing the strength of neutrino self-interactions for CMB and BAO data, although their discussion gives a more general description rather than for the case we study here. 

We numerically study changes to the phase shift in both the BAO oscillation signal and the CMB power spectra (TT, EE, TE) in the simplistic model for a four-fermion contact interaction. We explore this when $N_{\mathrm{eff}}$ or $G_{\mathrm{eff}}$ is varied. To reiterate, the goal is to understand whether there is some ability to use a measurement of the phase shift in these spectra to constrain $G_{\mathrm{eff}}$ (or any non-standard model particles expected to induce similar imprints on the phase-shift).
To do this, we use the code \textsc{CLASS-PT} of \cite{camarena2023confronting}\footnote{https://github.com/davidcato/class-interacting-neutrinos-PT} which has implemented a four-fermion contact interaction between neutrinos in the \textsc{CLASS} \cite{blas2011cosmic} Boltzmann solver, and allows one to compute the matter power spectrum and CMB power spectra with the neutrino self-interaction strength varied via $G_{\mathrm{eff}}$. This is of interest because a detection of the phase shift allows for a detection of the C$\nu$B that is more robust to changes in other cosmological parameters; in refs.~\citep{baumann2019first, montefalcone2025free}, a template has been constructed for the phase shift scale dependence due to Standard Model neutrinos that is independent of cosmological parameters; the amplitude of the phase shift (and thus the template) is simply scaled with changes to $N_{\mathrm{eff}}$, allowing for one to infer the presence of the C$\nu$B, and constrain $N_{\mathrm{eff}}$. 

While we can use the broadband shape of the matter power spectrum or CMB power spectra to study neutrino self-interactions, the impacts of changing the neutrino free-streaming properties are very degenerate with $A_s$ and $n_s$ given the accessible scales in the data. This has led to various works attempting to measure $G_{\mathrm{eff}}$ from cosmological data suffering from bimodality in the constraints. However, it may be possible for the phase shift to assist in constraining $G_{\mathrm{eff}}$ in a manner that does not suffer from this degeneracy. Unsurprisingly, we show the effect of altering $G_{\mathrm{eff}}$ for fixed $N_{\mathrm{eff}}$ on the phase $\Delta \ell$ (or $\Delta k$) is similar to the effect seen in ref.~\cite{choi2018probing} on the phase shift when free-streaming is delayed. For neutrinos with more greatly delayed free-streaming (for larger $G_{\mathrm{eff}}$), there is a decreasing amplitude for the phase shift and a turnover scale emerges which may be related to the scale identified in ref.~\cite{choi2018probing} as the horizon corresponding to the delayed neutrino free-streaming. In section~\ref{sec::selfinteractiontemplate}, we fit a template to capture the phase shift specifically in the case we are varying $G_{\mathrm{eff}}$, building on the template developed for Standard Model neutrinos in refs.~\cite{baumann2016phases, baumann2019first} in the BAO signal, and by ref.~\cite{montefalcone2025free} in the CMB power spectra.

\section{The phase shift of Standard Model neutrinos}\label{sec::standardmodelphase}

Relativistic neutrinos move close to the speed of light $c$, while the BAO sound speed is $\sim c/\sqrt{3}$. The parameter $N_{\mathrm{eff}}$ quantifies the abundance of any radiation species with this free-streaming behavior. Particles that propagate faster than the baryon-photon sound speed prior to the baryon drag epoch induce a characteristic phase shift in the power spectra ($\propto \Delta k$ in matter and $\propto \Delta \ell$ in the CMB), producing a robust signature of the C$\nu$B \citep{bashinsky2004neutrino, baumann2016phases, green2020phase}. This shift arises because free-streaming particles alter the gravitational potential, introducing a changes on the phase of BAO oscillations. Ref.~\cite{bashinsky2004neutrino} showed that modifying the driving of a harmonic oscillator -- an analogy for BAO perturbations -- changes both the amplitude and phase of the oscillations. Since neutrinos move faster than the baryon-photon sound speed, they generate metric perturbations beyond the sound horizon, shifting the physical scale of sound waves that freeze at recombination toward larger scales (smaller $k$ and $\ell$). 

% Break here this to a new paragraph
On small scales, the amplitude of the phase shift $\Delta \phi$ is proportional to $\epsilon_{\nu} = \frac{N_{\mathrm{eff}}}{N_{\mathrm{eff}} + \alpha_{\nu}}$, where $\alpha_{\nu} = \frac{8}{7}\left(\frac{11}{4}\right)^{4/3}$. However, on large scales the amplitude of the phase shift approaches zero. The phase shift was first parameterized and measured in CMB data by \cite{follin2015first}, then subsequently in BAO data by \cite{baumann2019first, wallisch2019cosmological} and more recently in the DESI DR1 BAO data by \cite{whitford2024constraining} and CMB data by \cite{montefalcone2025free}. Updated constraints from DESI BAOs using the second DESI data release (DR2) can be found in \cite{elbers2025constraints}. The CMB constraints by \cite{montefalcone2025free} show the presence of a non-zero phase shift at a significance of $14\sigma$, while the updated BAO constraints in \cite{elbers2025constraints}, with the addition of priors from Planck data, find a non-zero detection at a significance of $6.9\sigma$. 

\subsection{Baryon acoustic oscillations}\label{sec::smneutrinosbaos}

For Standard Model neutrinos, the impact of neutrino free-streaming is to shift the Fourier-space BAO signal toward smaller $k$. Ref.~\cite{bashinsky2004neutrino} showed that the shift in the phase, $\Delta \phi = \Delta k r_s$, is $\Delta \phi \approx 0.1912 \pi \epsilon_{\nu} $ for large $k$. %, where $\epsilon_{\nu} = \frac{N_{\mathrm{eff}}}{\alpha_{\nu} + N_{\mathrm{eff}}}$, $\alpha_{\nu} = \frac{8}{7} (\frac{11}{4})^{4/3}$. 
An improved calculation is given by \cite{green2020phase}, however a template was numerically fit to the phase shift by \cite{baumann2018searching, baumann2019first}, giving
\begin{align}
     \Delta \phi & = \beta_{\phi}(N_{\mathrm{eff}}) f(k)  \\ \nonumber 
     & = \frac{\epsilon_{\nu}}{ \epsilon_{\nu,\mathrm{fid}}} \frac{\phi_{\infty}}{1 + (k_*/k)^{\xi}}.
\end{align}
where $\beta_{\phi} = \epsilon_{\nu}/\epsilon_{\nu,\mathrm{fid}}$, and $\epsilon_{\nu,\mathrm{fid}}$ is $\epsilon_{\nu}$ for some fiducial cosmology. This template provides an effective representation of the phase shift  that captures the most important features and is independent of the cosmological model. The amplitude is captured by $\beta_{\phi}$, which is relative to the choice of $N_{\mathrm{eff}}$ in a fiducial model (labeled by `fid'). The function $f(k)$ captures the scale dependence of the phase shift, with $\phi_{\infty} = 0.227$, $k_* = 0.0324h$ Mpc$^{-1}$, and $\xi = 0.872$. 

\begin{figure}
    \centering
    \includegraphics[width=\linewidth]{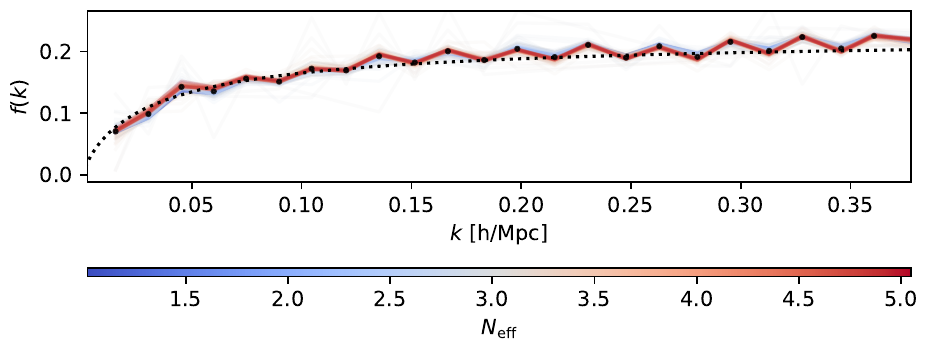}
    \caption{The phase shift function $f(k)$ extracted from the BAO signal. $f(k)$ has been extracted for 120 spectra by computing $\Delta k/ r_s$ relative to a spectrum with $N_{\mathrm{eff}} = 3.044$, for a range of values of $N_{\mathrm{eff}}$ indicated by the colour bar, with each normalized by $\beta_{\phi}(N_{\mathrm{eff}})$. The dotted line shows the fitting function given in \cite{baumann2018searching, baumann2019first}, and the black points show the average value of $f(k)$ from the 120 spectra. The power spectra were computed using \textsc{CLASS} \citep{blas2011cosmic}.\protect\footnote{https://github.com/lesgourg/class\_public}}
    \label{fig:BAO_signal_phase_standardmodelneutrinos}
\end{figure}

% Breaking here this to a new paragraph
%In order to see this shift in the BAO signal, 
To reveal the phase-shift signature of free-streaming radiation in the BAO signal independently of other cosmological effects, the oscillatory component must be extracted. 
%one can start by isolating the oscillatory BAO signal $\mathcal{O}(k)$, by smoothing the matter power spectrum, $P(k)$, to obtain $P_{\mathrm{sm}}(k)$, and computing $\mathcal{O}(k) = \frac{P(k)}{P_{\mathrm{sm}}(k)} - 1$. 
This is done by smoothing the matter power spectrum $P(k)$ to obtain $P_{\mathrm{sm}}(k)$, and then defining $\mathcal{O}(k) = \frac{P(k)}{P_{\mathrm{sm}}(k)} - 1$. Then, one can repeat this process for a range of power spectra with varying $N_{\mathrm{eff}}$. To ensure weak dependence to other effects, in each case the cold dark matter density needs to be chosen to fix the scale of matter-radiation equality without changing $\Omega_{\mathrm{b}}$, using $a_{\mathrm{eq}} = \frac{\Omega_{\mathrm{r}}}{\Omega_{\mathrm{m}}}$. We can set $\Omega_{\mathrm{cdm}}$ for a cosmology with $N_{\mathrm{eff,\mathrm{fid}}}$, $\Omega_{\mathrm{b,fid}}$, $\Omega_{\mathrm{cdm,fid}}$ as
\begin{equation}
    \Omega_{\mathrm{cdm}} = ( \frac{\alpha_{\nu} + N_{\mathrm{eff,\mathrm{fid}}}}{\alpha_{\nu} + N_{\mathrm{eff}} }) \Omega_{\mathrm{m}} - \Omega_{\mathrm{b}}.
\end{equation}
Furthermore, changes to the sound horizon scale need to be accounted for by rescaling $k$ for each of the power spectra by $\frac{r_s}{r_{s,\mathrm{fid}}}$, where $r_s$ can be computed by \textsc{CLASS} for each cosmology. Then, the shift in the peaks and troughs of the BAO oscillations $\Delta k$ can be compared to a fiducial model, and the average of the shifts in the peaks and troughs as a function of $k$ can be fit to give the function $f(k)$.\footnote{More detail on this algorithm can be found in \cite{baumann2018searching}.} The extracted phase shift signal from theoretical power spectra can be seen in Figure~\ref{fig:BAO_signal_phase_standardmodelneutrinos}.

\subsection{Cosmic microwave background}

\begin{figure}
    \centering
    \includegraphics[width=\linewidth]{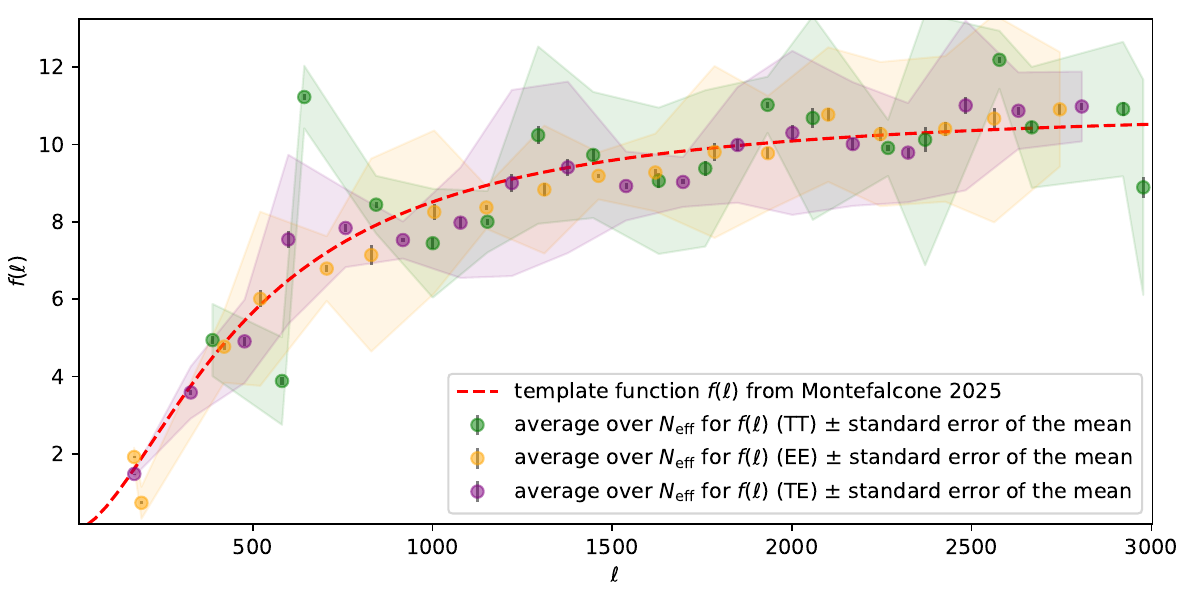}
    \caption{The scale dependent function $f_{\nu}(\ell)$ that captures the phase shift in the CMB power spectra, calculated by averaging the phase from 100 power spectra with varying $N_{\mathrm{eff}}$ after dividing by the amplitude $A_{\nu}$, defined in Equation~\eqref{eq:montefalconetemplate}. The red line shows the template of \cite{montefalcone2025free}, and the data points show the phase shift extracted from each of the spectra, with the shaded regions showing the spread in the averaged data points.  }
    \label{fig:phaseshiftneutrinos_CMB_standardmodel}
\end{figure}

The phase shift seen in the CMB power spectra has a functional form similar to that seen in BAO data. The functional form seen in the temperature (TT), polarization (EE), and cross temperature-polarization (TE) power spectra is well captured by the template fit in \cite{montefalcone2025free},
\begin{align}\label{eq:montefalconetemplate}
    \Delta \ell & = A_{\nu} f_{\nu} (\ell) \\ \nonumber 
    & = \frac{(\epsilon_{\nu} - \epsilon_{\nu,N_{\mathrm{eff}} = 3.044})}{(\epsilon_{\nu,N_{\mathrm{eff}} = 1.0} - \epsilon_{\nu,N_{\mathrm{eff}} = 3.044})} \frac{\ell_{\infty}}{1 + (\ell/\ell_{*})^{\xi}}.
\end{align}
The phase shift is given by $\Delta \phi = \Delta \ell \theta_s$, where $\theta_s$ is the sound horizon at recombination. The amplitude of the phase shift $A_{\nu} = (\epsilon_{\nu} - \epsilon_{\nu,N_{\mathrm{eff}} = 3.044})/(\epsilon_{\nu,N_{\mathrm{eff}} = 1.0} - \epsilon_{\nu,N_{\mathrm{eff}} = 3.044})$ is normalized with respect to the difference in the amplitude of the shift for $N_{\mathrm{eff}} = 3.044$ relative to $N_{\mathrm{eff}} = 1.0$. The function $f_{\nu}(\ell)$ is shown in Figure~\ref{fig:phaseshiftneutrinos_CMB_standardmodel}, along with the phase shift extracted from the CMB power spectra using an approach similar to the procedure used for the BAO signal; this is also described in detail in \cite{follin2015first, hou2013massless, montefalcone2025free}. While the choice of normalization is arbitrary, we would like to express the normalization in a manner similar to that for the BAO; we capture the change in the phase shift amplitude with the parameter $\beta_{\phi} = \frac{\epsilon_{\nu}}{\epsilon_{\nu,\mathrm{fid}}}$, which is unity when the fiducial cosmological model and the true cosmology match. Therefore we equivalently write Equation~\eqref{eq:montefalconetemplate} as

\begin{align}
    \Delta \ell &= \frac{\epsilon_{\nu,N_{\mathrm{eff}} = 3.044}}{(\epsilon_{\nu,N_{\mathrm{eff}} = 1.0} - \epsilon_{\nu,N_{\mathrm{eff}} = 3.044})} \frac{(\epsilon_{\nu} - \epsilon_{\nu,N_{\mathrm{eff}} = 3.044})}{\epsilon_{\nu,N_{\mathrm{eff}} = 3.044}} f_{\nu}(\ell) \\ \nonumber 
    &= \left( \frac{\epsilon_{\nu,N_{\mathrm{eff}} = 3.044}}{\epsilon_{\nu,N_{\mathrm{eff}} = 1.0} - \epsilon_{\nu,N_{\mathrm{eff}} = 3.044}} \right) [\beta_{\phi} - 1] f_{\nu}(\ell).
\end{align}
This parametrizes the phase shift amplitude in terms of $\beta_{\phi}$, analagous to the BAO phase shift description, and the factor of $\epsilon_{\nu,N_{\mathrm{eff}} = 3.044}/(\epsilon_{\nu,N_{\mathrm{eff}} = 1.0} - \epsilon_{\nu,N_{\mathrm{eff}} = 3.044})$ is simply a constant factor. This choice specifically forces the fiducial cosmology to correspond to $N_{\mathrm{eff}} = 3.044$ in the denominator of $\beta_{\phi}$.\footnote{One also needs to account for the fact that in this writing of the phase shift $\Delta \ell$ is actually negative if $N_{\mathrm{eff}} > 1.0$ (for the true cosmology), because the constant factor $\epsilon_{\nu,N_{\mathrm{eff}} = 3.044}/(\epsilon_{\nu,N_{\mathrm{eff}} = 1.0} - \epsilon_{\nu,N_{\mathrm{eff}} = 3.044})$ in front of the equation is negative. However, the phase shift $\Delta \phi$ in a cosmology with a positive number of neutrino species, that shifts $\mathcal{C}(\ell) \rightarrow \mathrm{C}(\ell + \Delta \phi)$, relative to a cosmology with $N_{\mathrm{eff}} = 0$, should shift the peaks to smaller values of $\ell$, which is only true is $\Delta \phi$ is positive. Therefore, the above equation acquires a negative sign above, unless one directly maps $\mathcal{C}(\ell)$ evaluated at $\ell$, to $\ell + \Delta \phi$, rather than mapping $\ell$ to $\mathcal{C}(\ell + \Delta \phi)$.}

As for the BAO, in order to isolate the phase shift signature independently of other cosmological parameters, one needs to alter $\Omega_{\mathrm{cdm}}$ to fix the epoch of matter-radiation equality as $N_{\mathrm{eff}}$ is varied, and the change in scale $\theta_s$ (computed by CLASS) is used to rescale the $\ell$'s to fix the sound horizon at the epoch of recombination. In the CMB, one must also fix the diffusion damping scale $r_D$, which can be controlled by $Y_p$, the helium abundance (mass fraction). We fix the ratio $r_D/r_s$ using the approximate relation \citep{hou2013massless}
\begin{equation}
    \frac{r_D}{r_s} = \frac{\theta_D}{\theta_s} \approx \frac{(1 + \rho_{\nu}/\rho_{\gamma})^{0.28}}{\sqrt{1 - Y_p}}.
\end{equation}
The ratio $\rho_{\nu}/\rho_{\gamma}$ can be computed as $N_{\mathrm{eff}} \alpha_{\nu}^{-1}$. The power of 0.28, seen in the numerator, may not be optimal for all cosmological models, but we found this choice worked sufficiently well for our choice of fiducial cosmology. We note the oscillations seen in both Figure~\ref{fig:BAO_signal_phase_standardmodelneutrinos} and Figure~\ref{fig:phaseshiftneutrinos_CMB_standardmodel} contain both physical and noise contributions, with the physical components arising since the phase shift template is only a leading order approximation; nonetheless, the templates capture the important features we can reasonably hope to constrain.

\section{The phase shift with non-standard model neutrino physics}\label{sec::selfinteractiontemplate}

Now we explore how the phase shift changes in a power spectrum generated for a cosmology with non-standard model neutrinos. As mentioned before, we consider the case of self-interacting neutrinos following a four-fermion interaction with universal coupling, whose effective strength is parameterized by $G_{\mathrm{eff}}$. We apply the same algorithms as discussed above for both the BAO and the CMB to see the phase shift, but now for the case in which $G_{\mathrm{eff}}$ is varied between $[-6, 0.5]$. For very small $G_{\mathrm{eff}}$, we expect the power spectrum to match the case of $\Lambda$CDM with Standard Model neutrinos. This should recover the phase shift signal seen in the previous sections, with the amplitude modulated by the value of $N_{\mathrm{eff}}$, as demonstrated in Appendix~\ref{App:B}. Like the phase shift introduced by Standard Model neutrinos, we also expect the impact of $G_{\mathrm{eff}}$ on the phase shift to be roughly independent of $z$ as the physics that causes the phase shift in the early Universe remains unaffected by the physics of the late Universe.

In practice, we fit multiple templates to the phase shift signal by varying the power spectra with $N_{\mathrm{eff}}$ to obtain a function analogous to $f(k)$, for various fixed choices of $G_{\mathrm{eff}}$.\footnote{It may also be possible to alternatively fix $N_{\mathrm{eff}}$, and instead fit a function analogous to $f(k)$ due to changing $G_{\mathrm{eff}}$, and repeat this for different values of $N_{\mathrm{eff}}$. This is illustrated in Figure~\ref{fig:phaseshiftdemo_BAO}. However, we found fitting a template to this to be much more difficult due to the non-linear impact of $G_{\mathrm{eff}}$. } This approach allows one to easily obtain the averaged non-linear dependence of the phase shift for different fixed values of $G_{\mathrm{eff}}$. $G_{\mathrm{eff}}$ changes the functional form of the phase shift due to delayed free-streaming. Like the algorithm described previously, we compute the shift $\Delta k$ for many values of $N_{\mathrm{eff}}$ while dividing out the change in amplitude $\beta_{\phi}$. The main impact of varying $N_{\mathrm{eff}}$ is simply to linearly rescale the amplitude of the phase shift. This is expected to be the case regardless of the value of $G_{\mathrm{eff}}$ since the neutrino interactions are universal in the model we use; for alternative cases where interactions are not universal, it would be necessary to study the phase shift that arises in these circumstances separately. While the model studied here and variations of self-interaction models already very strongly constrained by terrestrial experiments \citep{lyu2021self, blinov2019constraining, belotsky2001constraint, berkov1987possible, berkov1988possible}, it allows us to test and understand whether there is information available regarding non-standard interactions in the phase shift of the power spectra that could be accessible with future experiments.  

\subsection{Baryon Acoustic Oscillations}

\begin{figure}[h!]
    \begin{subfigure}{0.5\textwidth}
    \includegraphics[width=\linewidth]{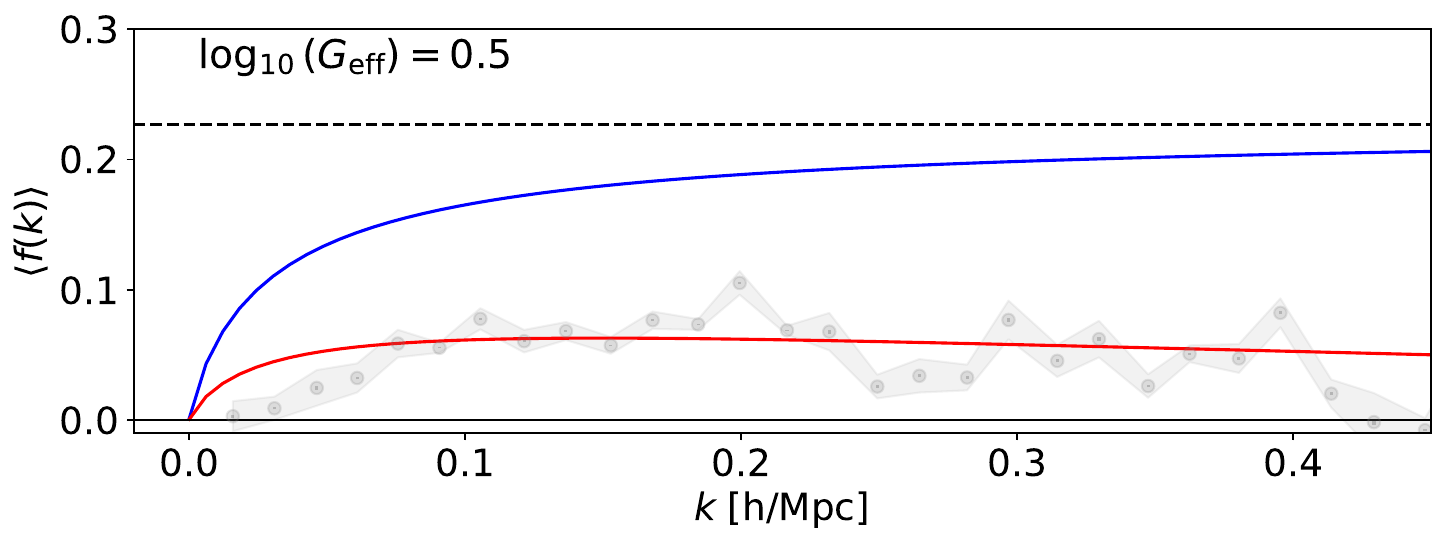}
    \end{subfigure}\begin{subfigure}{0.5\textwidth}
    \includegraphics[width=\linewidth]{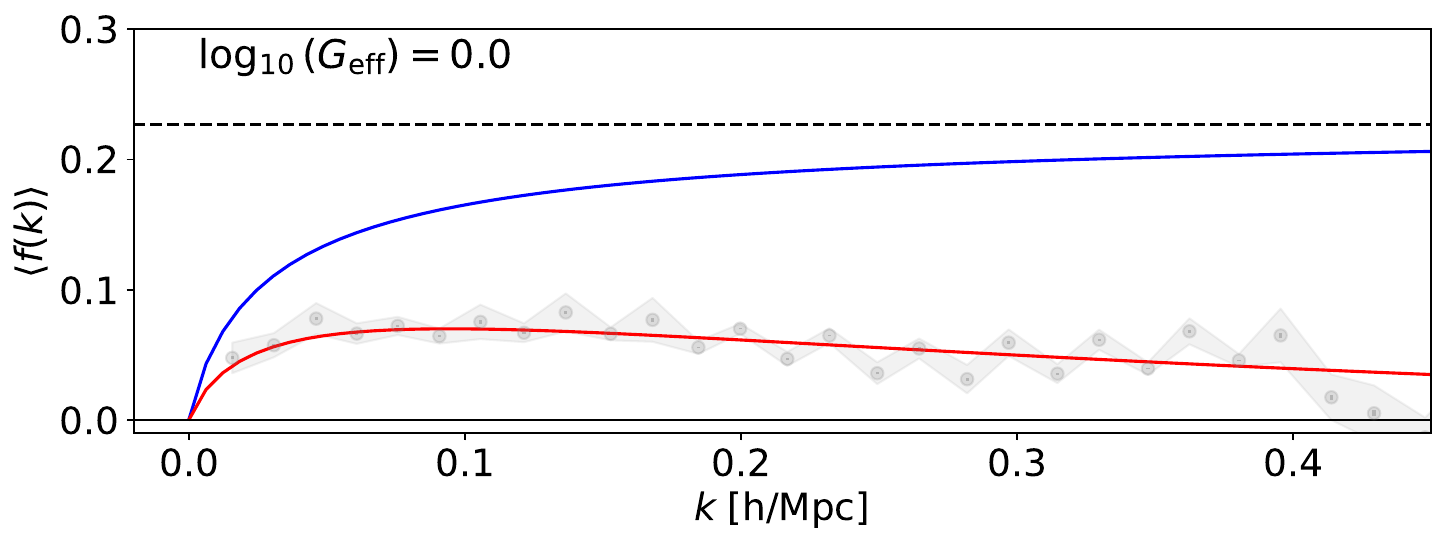}
    \end{subfigure}
    
    \begin{subfigure}{0.5\textwidth}
    \includegraphics[width=\linewidth]{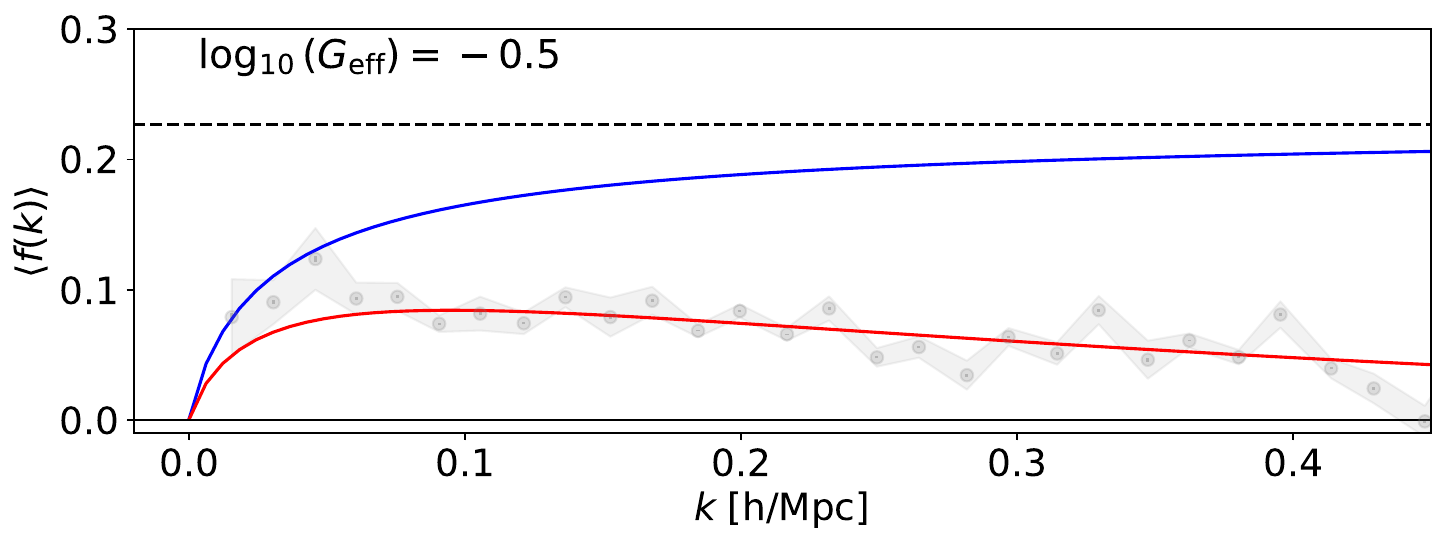}
    \end{subfigure}\begin{subfigure}{0.5\textwidth}
    \includegraphics[width=\linewidth]{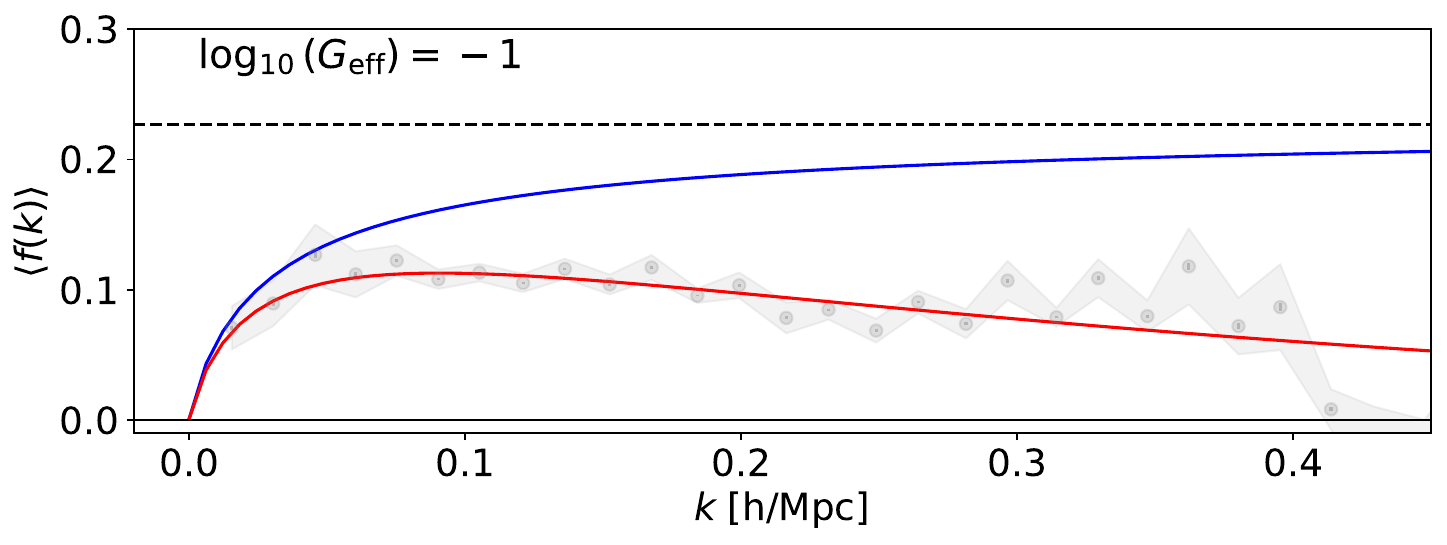}
    \end{subfigure}
    
    \begin{subfigure}{0.5\textwidth}
    \includegraphics[width=\linewidth]{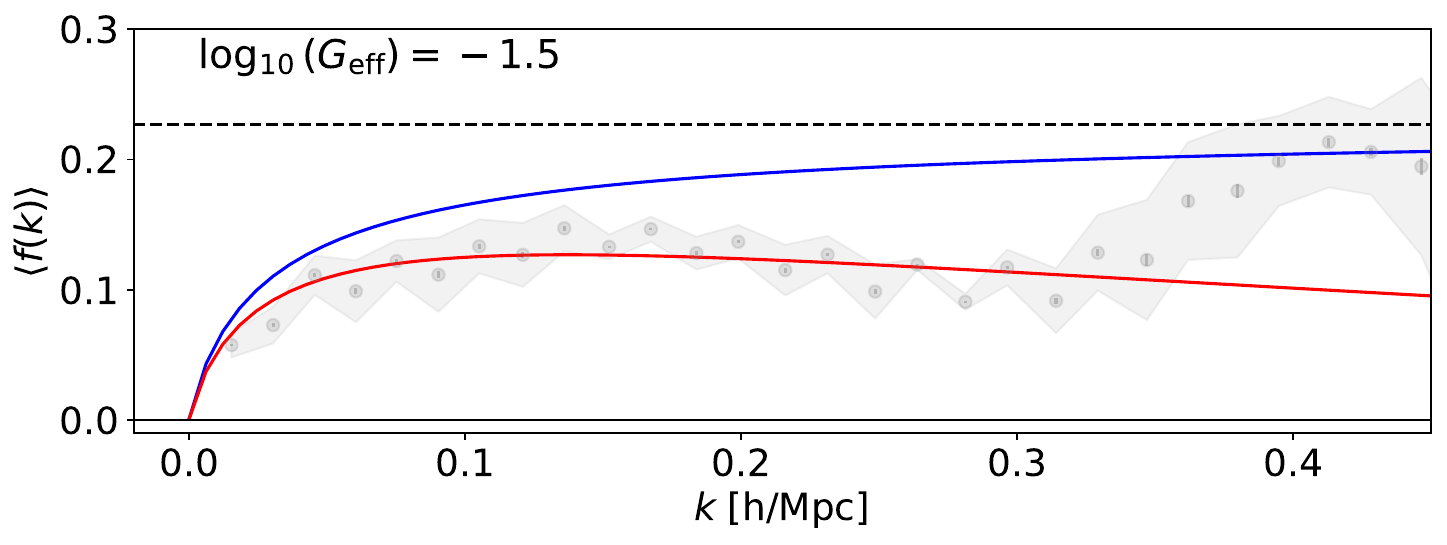}
    %\caption{$\log_{10}{(G_{\mathrm{eff}})} = -1.5.$}
    \end{subfigure}\begin{subfigure}{0.5\textwidth}
    \includegraphics[width=\linewidth]{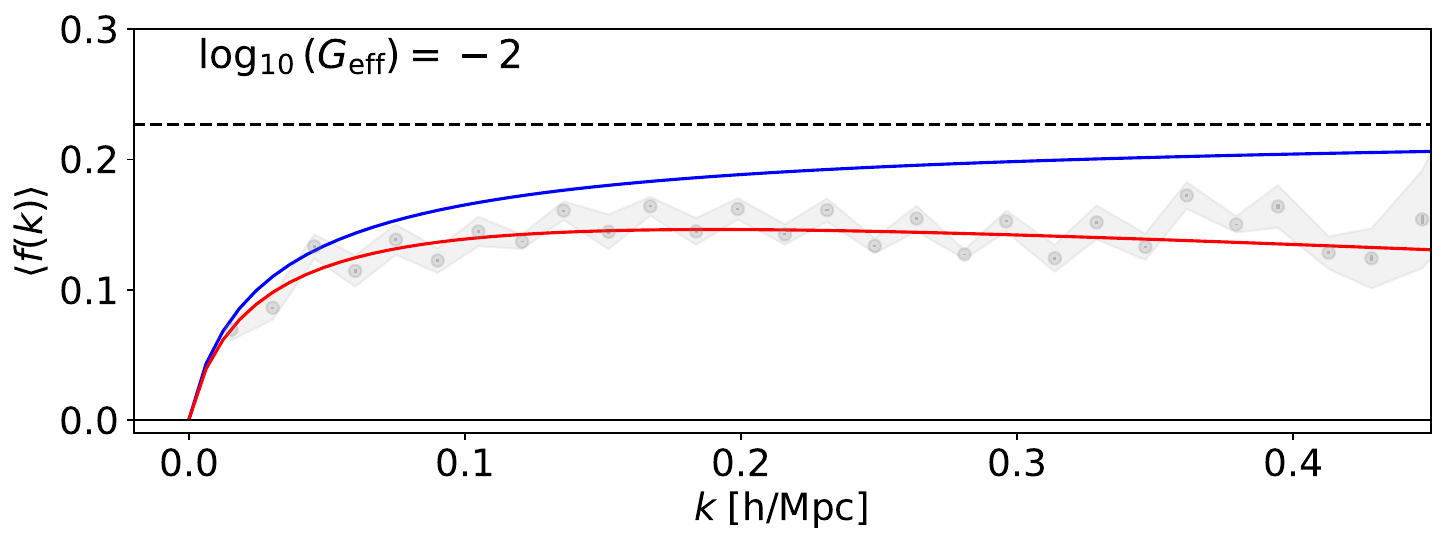}
    %\caption{$\log_{10}{(G_{\mathrm{eff}})} = -2.$}
    \end{subfigure}
    
    \begin{subfigure}{0.5\textwidth}
    \includegraphics[width=\linewidth]{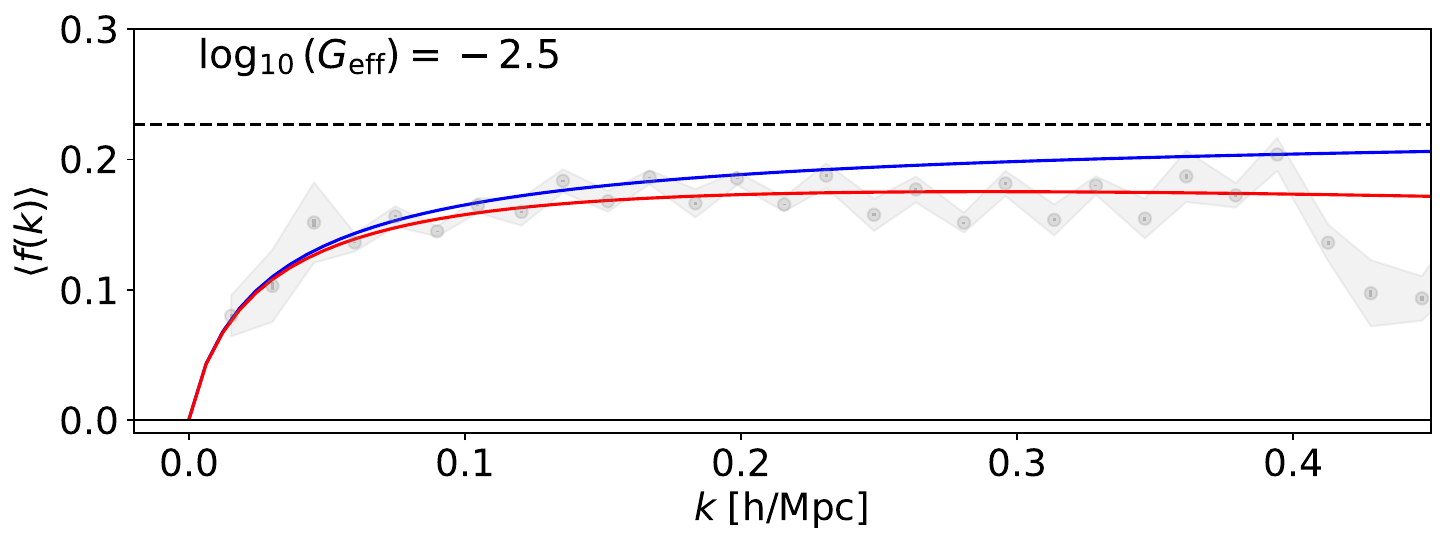}
    %\caption{$\log_{10}{(G_{\mathrm{eff}})} = -2.5$}
    \end{subfigure}\begin{subfigure}{0.5\textwidth}
    \includegraphics[width=\linewidth]{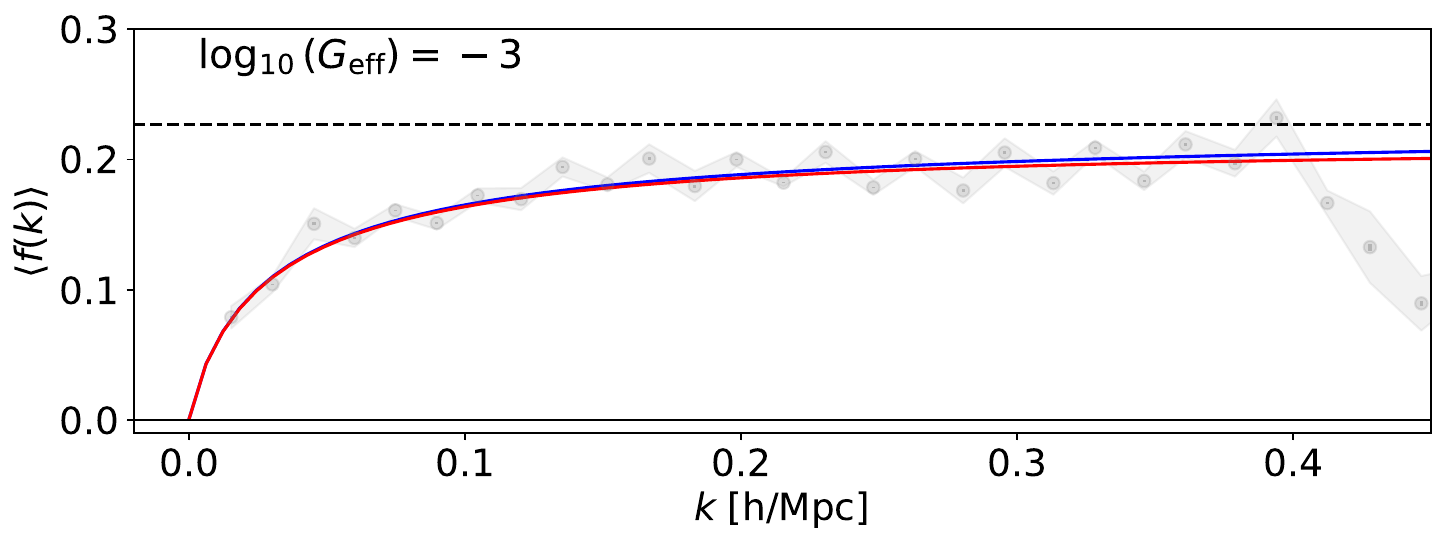}
    %\caption{$\log_{10}{(G_{\mathrm{eff}})} = -3$}
    \end{subfigure}
    
    \begin{subfigure}{0.5\textwidth}
    \includegraphics[width=\linewidth]{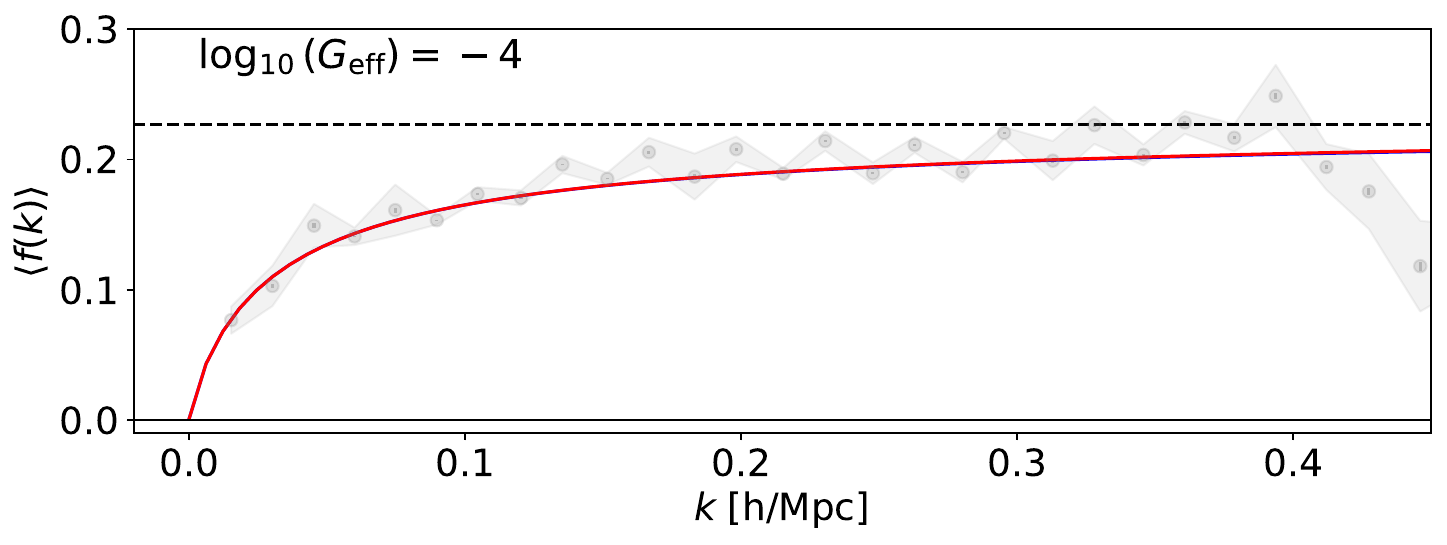}
    %\caption{$\log_{10}{(G_{\mathrm{eff}})} = -4$}
    \end{subfigure}\begin{subfigure}{0.5\textwidth}
    \includegraphics[width=\linewidth]{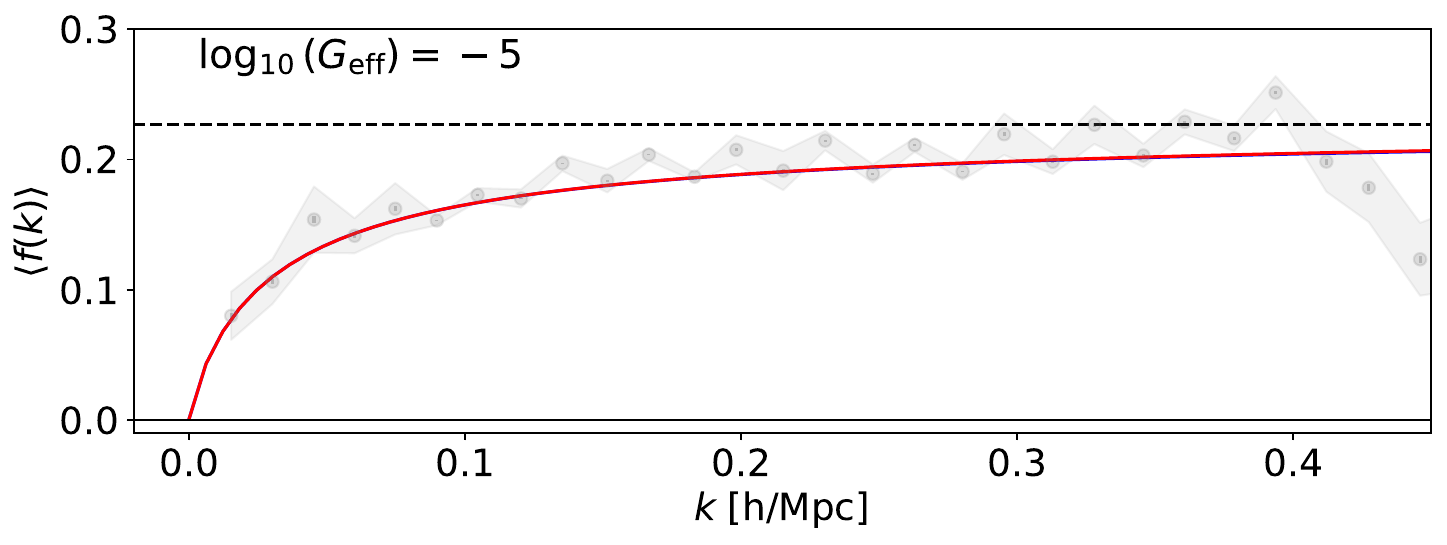}
    %\caption{$\log_{10}{(G_{\mathrm{eff}})} = -6$}
    \end{subfigure}
    \caption{The phase shift signal for different fixed values of $G_{\mathrm{eff}}$. The red line shows a fit through the extracted phase shift and the blue line shows the template for Standard Model neutrinos \citep{baumann2018searching, wallisch2019cosmological}. The dashed black lines correspond to the standard model theoretical prediction derived by \cite{bashinsky2004neutrino}. The numerical calculations are noisy for $ -2 < \log_{10}{(G_{\mathrm{eff}})} < -1$, likely because there is more power on scales close to the BAO scale for these interaction strengths which introduces noise into the dewiggling process. Additionally, for all plots on scales $> 0.4 h\, \mathrm{Mpc}^{-1}$ the BAO oscillations are very small, leading to a noisier signal. As such we fit only for $k \leq 0.4 h\, \mathrm{Mpc}^{-1}$. The power spectra were generated using \textsc{CLASS-PT}. }\label{fig:templatefitsGeff}
\end{figure}

We apply the same algorithm as discussed in section~\ref{sec::smneutrinosbaos} for the BAO signal, but now multiple times for $G_{\mathrm{eff}}$ varied between $[-6, 0.5]$. As we expect, we find that for very small $G_{\mathrm{eff}}$ the power spectrum and phase shift is effectively the same as for the case of $\Lambda$CDM with Standard Model neutrinos. In Figure~\ref{fig:templatefitsGeff} we show the numerically extracted phase shift with varying $G_{\mathrm{eff}}$, and a fit through it. % For the fits, we have simply taken the function $f(k)$ of \cite{baumann2019first, wallisch2019cosmological} and allow it to have a varying amplitude $A$ and an additional exponential damping controlled by $B$, both of which are functions of $G_{\mathrm{eff}}$,
In practice, we fit the phase shift considering the functional form of $f(k)$ for the Standard Model case \citep{baumann2019first, wallisch2019cosmological} modulated by extra variations on amplitude and scale-dependence due to the delayed free streaming. This contrasts with the fits shown in \cite{montefalcone2025directly} for the CMB, in which they fit only the amplitude change and neglect the scale dependent change in the phase shift signal. However, carefully studying Figures 2, 6 and 7 in their work shows a similar scale-dependent change in the signal. In line with the scale-dependent damping and shifting `peak' seen in the results of \cite{choi2018probing}, we find that such changes can be well-represented by a changes in amplitude $A$ and additional exponential damping parameterized by $B$, both functions of $G_{\mathrm{eff}}$, such we have
\begin{equation}\label{eq:templatelog10geff_BAO}
    \Delta \phi = A(G_{\mathrm{eff}}) f(k) e^{k r_s B(G_{\mathrm{eff}})}. 
\end{equation}
In each case, the parameters $A$ and $B$ have been fit using the package \textsc{emcee} \citep{foreman2013emcee},\footnote{https://emcee.readthedocs.io/en/stable/\#} and this equation and the figure shows the phase shift amplitude for $N_{\mathrm{eff}} = 3.044$ relative to a spectrum with $N_{\mathrm{eff}} = 0$. We found that using \textsc{emcee} gave results fairly consistent with using the \textsc{scipy.optimize.minimize} algorithm \citep{virtanen2020scipy, 2020SciPy-NMeth, mckinney-proc-scipy-2010} but also allowed us to estimate uncertainties on $A$ and $B$. More generally, we could write the shift in the $k$s in the matter power spectrum oscillations relative to $\Lambda$CDM with $N_{\mathrm{eff}} = 3.044$ as 
\begin{equation}
    \mathcal{O}(k) \rightarrow \mathcal{O}(k + [A(G_{\mathrm{eff}})\beta_{\phi}(N_{\mathrm{eff}}) e^{k B(G_{\mathrm{eff}}) r_s} - 1] f(k)/r_s).  
\end{equation}
In the limit that $G_{\mathrm{eff}} \rightarrow 0$, we have the standard model phase shift as $A \rightarrow 1$ and $B \rightarrow 0$, giving 
\begin{equation}
    \mathcal{O}(k) \rightarrow \mathcal{O}(k + [\beta_{\phi}(N_{\mathrm{eff}}) - 1] f(k)/r_s), 
\end{equation}
which in turn reduces to the standard BAO wiggles in the case where $N_{\mathrm{eff}}$ is equal to the fiducial value.

\begin{figure}
    \centering
    \includegraphics[width=\linewidth]{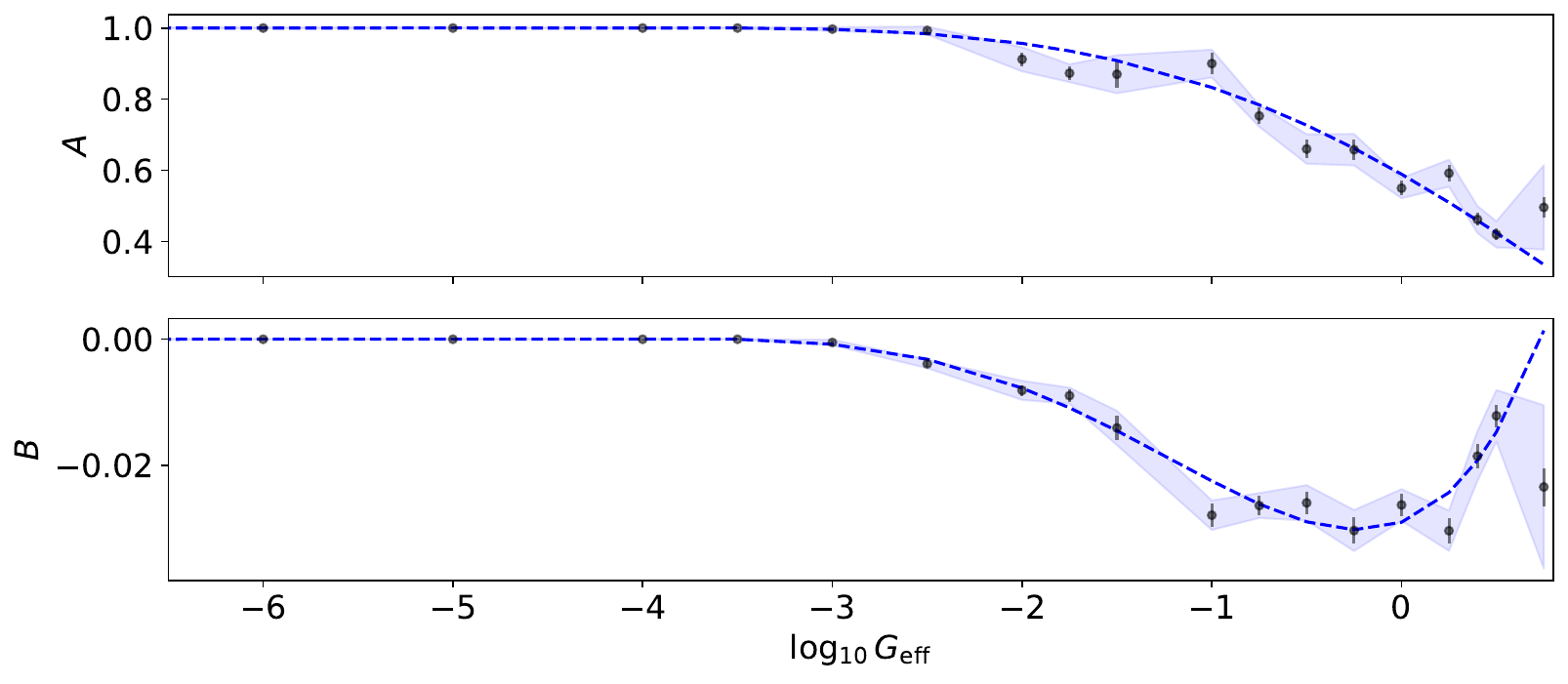}
    \caption{The fit to the phase-shift amplitude and exponential damping parameters $A$ and $B$ in Equation~\eqref{eq:templatelog10geff_BAO} as a function of $\log_{10}{(G_{\mathrm{eff}})}$. We use a 6th order polynomial to fit a smooth function through the points extracted from the phase shift calculations. The points in the fit for the template as a function of $k$ has been weighted by the uncertainty on the individual fits for each parameter $A$ and $B$. The error bars have been scaled by the $\chi^2$ goodness-of-fit to each point.  }
    \label{fig:templatefitlog10Geff_BAO}
\end{figure}

In Figure~\ref{fig:templatefitlog10Geff_BAO} we show the fit to the parameters $A$ and $B$ as a function of $\log_{10}{(G_{\mathrm{eff}})}$.\footnote{We employ 6th-order polynomial fits to represent $A$ and $B$ as functions of $\log_{10}{(G_{\mathrm{eff}})}$. We have found this particular choice to be the most suitable for our Forecast analysis. Other approaches, as a rational construction of these functions, might outperform our choice when applied to real data.} The change to the phase shift signal with varying $\log_{10}{(G_{\mathrm{eff}})}$ is well captured by the parameter $A$ that modulates the amplitude of the signal, and $B$ that modulates an exponential damping. As shown by the uppermost panel of Figure~\ref{fig:templatefitsGeff}, the fit appears less accurate for very strong self-interactions. Although improved modeling could be explored in future work, the discrepancy is likely due to numerical noise, since the phase-shift signal is suppressed at strong interaction strengths.

The variation of the amplitude $A$ with $G_{\mathrm{eff}}$ is expected to be at least partially degenerate with the effect of $N_{\mathrm{eff}}$, which also modulates the phase-shift amplitude through $\beta_{\phi}$. This degeneracy will weaken the constraints on $G_{\mathrm{eff}}$. Given that the BAO signal only puts weak constraints on $N_{\mathrm{eff}}$ alone from the phase shift at present, this suggests that at present it is unlikely that the BAO will be able to strongly constrain $G_{\mathrm{eff}}$ from just the phase shift signal alone (although we quantify this more clearly in Section~\ref{sec::constraintsBAO}). However, it may be possible to combine BAO and CMB information to obtain better constraints. Additionally, the signal is quite subtle: it does not change quickly for $\log_{10}{(G_{\mathrm{eff}})} \geq -0.5$; while for $\log_{10}{(G_{\mathrm{eff}})} \leq -3.5$ there is no discernible impact on the phase shift at all. This is also inline with the results presented by \cite{montefalcone2025directly} for the CMB, who find the phase shift approaches a constant in the limit of fluid-like neutrinos (corresponding to stronger interactions). For weaker interaction strengths, neutrino free-streaming is not significantly delayed.

We note that we encountered difficulty in dewiggling the power spectrum and thus extracting the phase shift signal for $ -2 \leq \log_{10}{(G_{\mathrm{eff}})} \leq -1$. The power spectrum wiggles and smooth broadband signal could not be robustly separated by simply applying the dewiggling algorithm of \cite{hinton2016measuring} or \cite{wallisch2019cosmological}. We identified this as being caused by the additional power that the neutrino self-interactions leave in the full shape matter power spectrum due to the delayed free-streaming; see the second panel in Figure~\ref{fig:phaseshiftdemo_BAO}. Relative to the $\Lambda$CDM power spectrum, there is a `bump' on scales close to the BAO scale for $ -2 \leq \log_{10}{(G_{\mathrm{eff}})} \leq -1$, which possibly makes it more difficult to fit a smooth polynomial through the spectrum (whether in real space or Fourier space). However, we found the best approach to improve the robustness of these methodologies was to simply vary $A_s$ and $n_s$ in a manner that partially cancels the additional power that enters the spectrum on those scales. This allowed the dewiggling algorithm of \citep{wallisch2019cosmological} to work more robustly. This solution in altering $A_s$ and $n_s$ does not impact the phase shift of the BAO oscillations since $A_s$ and $n_s$ have no impact on the sound horizon scale or positions of the peaks. We used the following equations to vary $A_s$ and $n_s$,

\begin{align}\label{eq:As_ns_equations}
    \ln{(10^{10}A_s)} &= -0.28\log_{10}{G_{\mathrm{eff}}} + 2.62, \\ \nonumber 
    n_s &= -0.076\log_{10}{G_{\mathrm{eff}}} + 0.835.
\end{align}

\subsection{Cosmic microwave background}

\begin{figure}
    \begin{subfigure}{0.5\textwidth} \includegraphics[width=\linewidth]{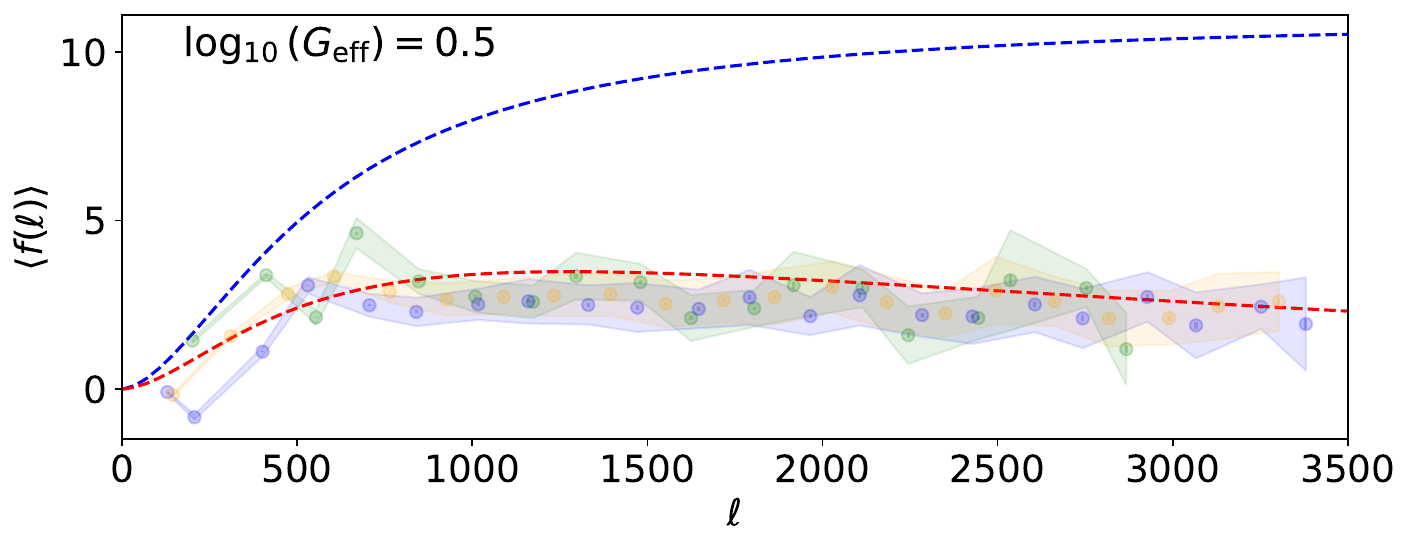}
    \end{subfigure}\begin{subfigure}{0.5\textwidth} \includegraphics[width=\linewidth]{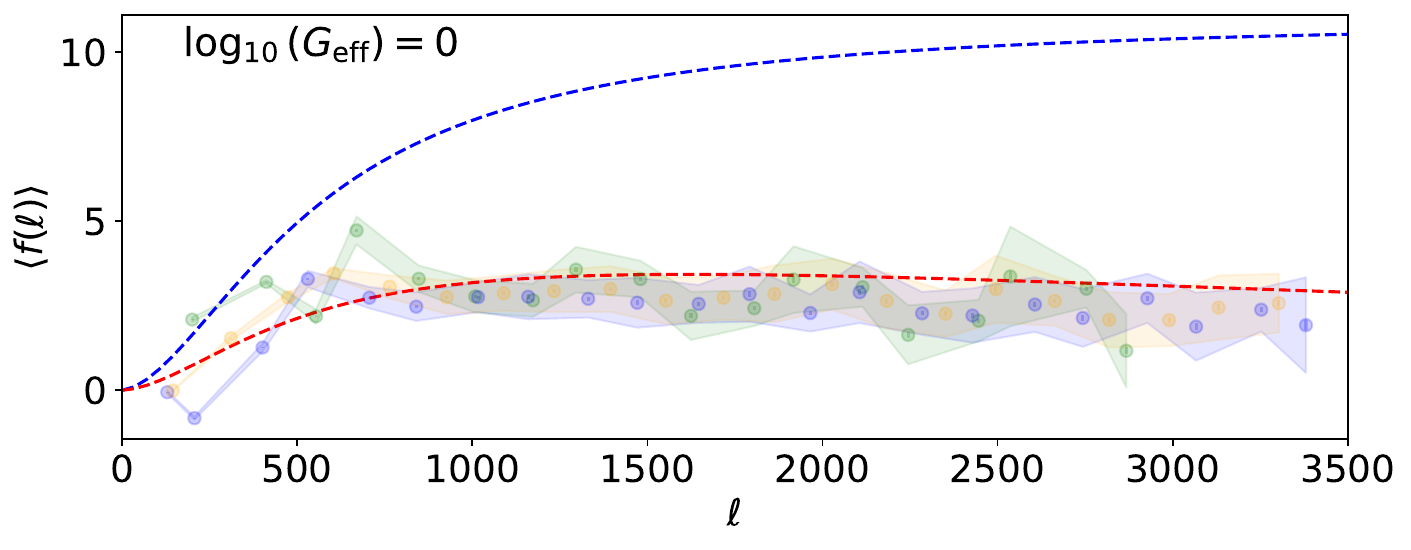}
    \end{subfigure}
    \begin{subfigure}{0.5\textwidth} \includegraphics[width=\linewidth]{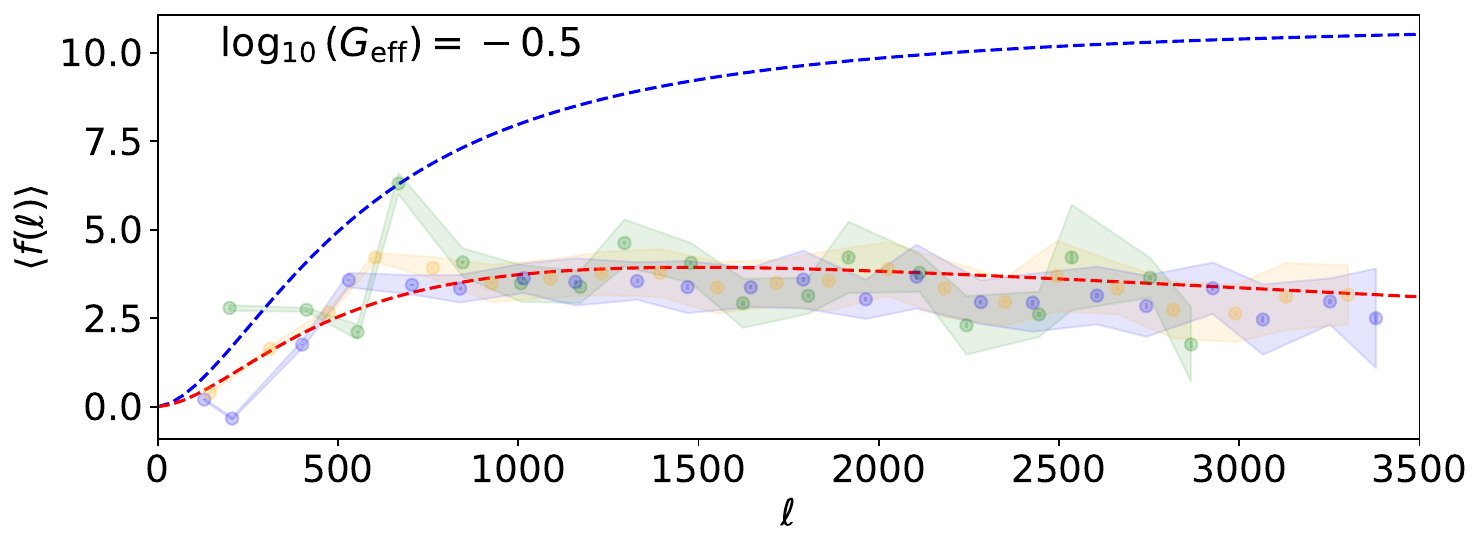}
    \end{subfigure}
    \begin{subfigure}{0.5\textwidth} \includegraphics[width=\linewidth]{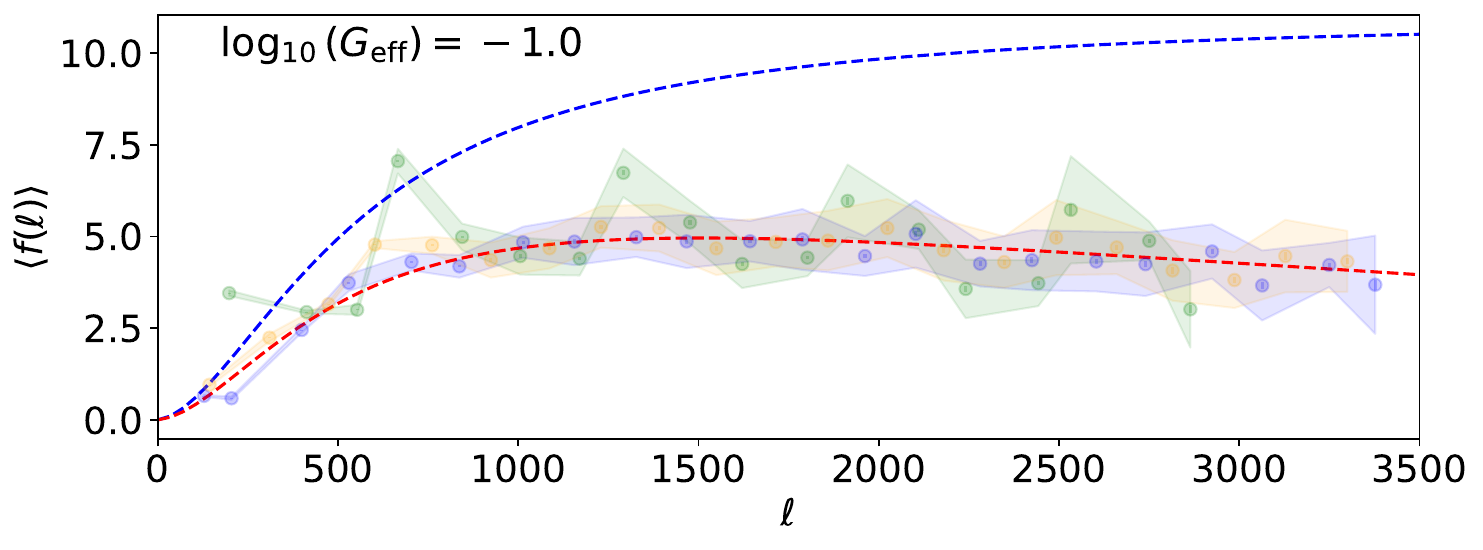}
    \end{subfigure}
    \begin{subfigure}{0.5\textwidth} \includegraphics[width=\linewidth]{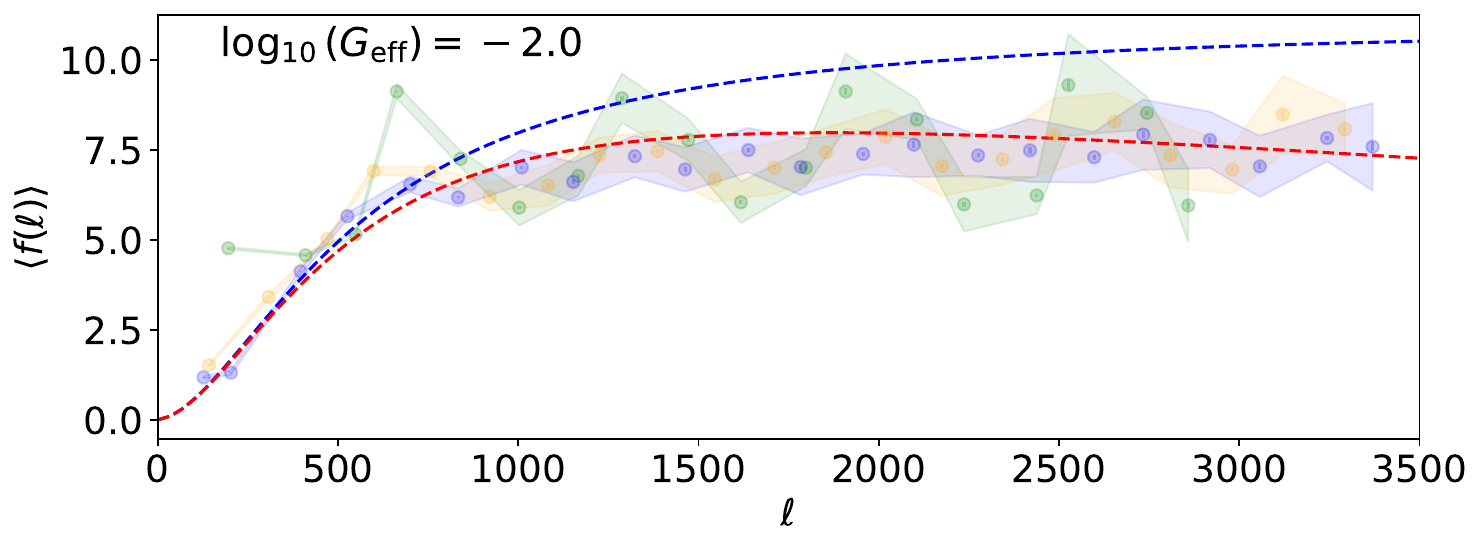}
    \end{subfigure}
    \begin{subfigure}{0.5\textwidth} \includegraphics[width=\linewidth]{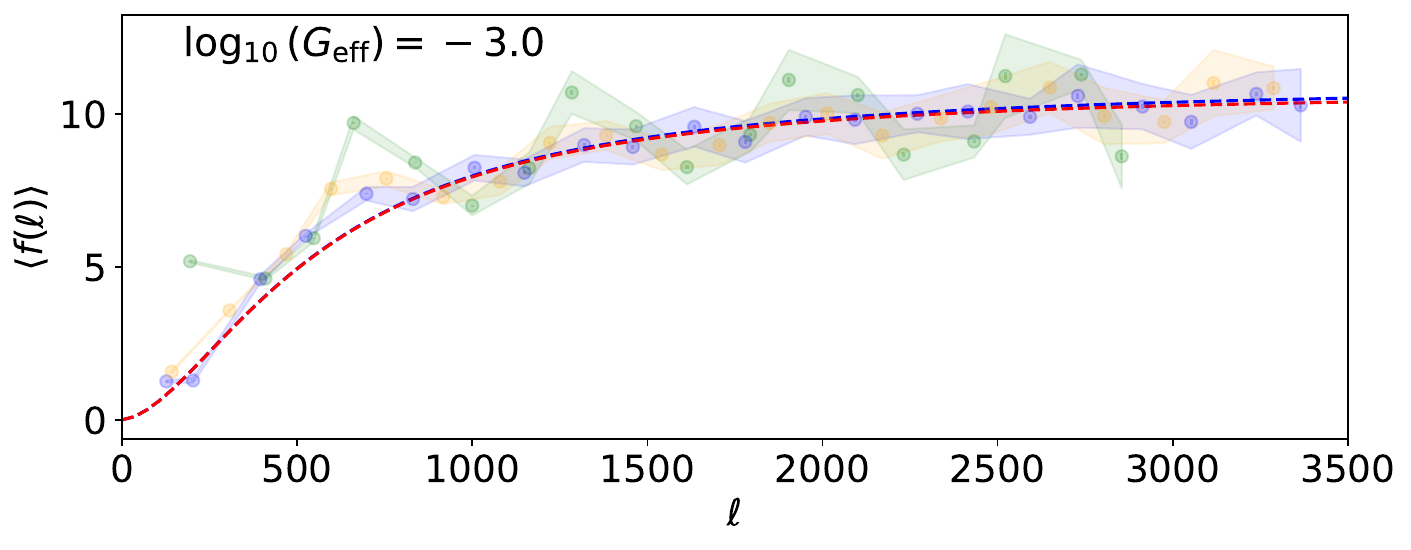}
    \end{subfigure}
    \caption{The phase shift signal for different fixed values of $G_{\mathrm{eff}}$ in the CMB power spectra. The red line shows a fit through the extracted phase shift and the blue line shows the template for Standard Model neutrinos \citep{montefalcone2025free}. The green, blue and yellow regions show the shift in $\ell$ for the TT, TE and EE power spectra respectively. The power spectra were generated using \textsc{CLASS-PT}. Correlations between each of the power spectra have been neglected in these fits, but we include cross-correlations  later for estimating the constraining power of the spectra for $\log_{10}{(G_{\mathrm{eff}})}$.}\label{fig:templatefitting_CMB}
\end{figure}

\begin{figure}
    \centering
    \includegraphics[width=\linewidth]{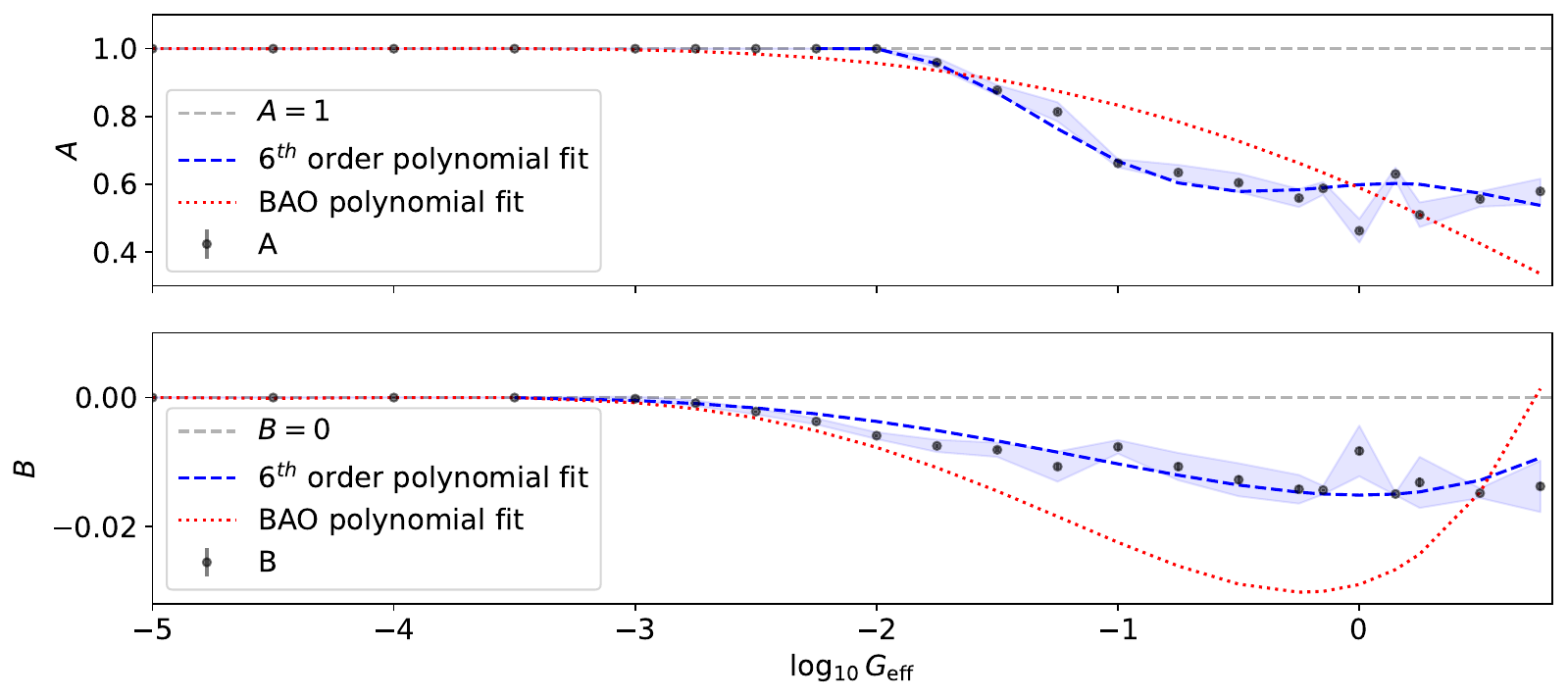}
    \caption{The fit to $A$ and $B$ as a function of $G_{\mathrm{eff}}$.  We recover $A = 1$ and $B = 0$ at very negative values of $\log_{10}{(G_{\mathrm{eff}})}$, where we expect the power spectrum and phase shift from this model to agree with the expectation of $\Lambda$CDM. In each panel, the blue dashed line shows a polynomial fit through the parameters; for smaller $G_{\mathrm{eff}}$ we fit a simple weighted polynomial, as for the BAO, but we only fit a subset of the points (shown by the blue dashed line) and model $A = 1$ and $B = 0$ as constants for larger values of $G_{\mathrm{eff}}$. Some of the noise in the fit to $A$ and $B$ is likely due to the degeneracy between these parameters. For comparison the BAO polynomial fits are also shown by the red dotted lines.}
    \label{fig:fittingparameters_CMB_geff}
\end{figure}

As before for the BAO signal, we explore how the phase shift signal changes in the CMB power spectra for neutrinos with non-standard interactions by varying $G_{\mathrm{eff}}$. Figure~\ref{fig:templatefitting_CMB} shows how the phase shift signal is impacted by self-interactions, and a model fit through the modified phase shift signal. The phase shift can be seen very clearly in the CMB data; we refit the parameters $A$ and $B$ as before, keeping the template $f_{\nu}(\ell)$ from Equation~\eqref{eq:montefalconetemplate} and modulating the amplitude via $A$ and an exponential damping through $B$ for each value of $\log_{10}{(G_{\mathrm{eff}})}$;
\begin{equation}
    \Delta \ell = A(G_{\mathrm{eff}}) f_{\nu}(\ell) e^{(B (G_{\mathrm{eff}})/50 \theta_s \ell) }.
\end{equation}
Similar to our fits for the BAO, we find the parameter $B$ captures a small but significant scale-dependent change to the phase shift amplitude, which is neglected in \cite{montefalcone2025directly} despite being visible in some of their figures. In Figure~\ref{fig:fittingparameters_CMB_geff} we show the fit to these parameters as a function of $G_{\mathrm{eff}}$. As before, we use \textsc{emcee} to fit the parameters. The function $f_{\nu}(\ell)$ has been defined in \cite{montefalcone2025free} to be the phase shift for $N_{\mathrm{eff}} = 3.044$ relative to $N_{\mathrm{eff}} = 1.0$. As such we can express the shift in the $\ell$s more generally to allow one to see the shift relative to an arbitrary choice of $N_{\mathrm{eff}}$ if desired,
\begin{equation}
    C_{\ell}(\ell) \rightarrow C_{\ell}(\ell + [\beta(N_{\mathrm{eff}}) A(G_{\mathrm{eff}}) e^{(B (G_{\mathrm{eff}})/50 \theta_s \ell)} - 1] C_{\nu} f_{\nu}(\ell) ). 
\end{equation}
Here $C_{\nu}$ is simply a constant equal to $\frac{-\epsilon_{\nu,N_{\mathrm{eff}} = 3.044}}{(\epsilon_{\nu,N_{\mathrm{eff}} = 1.0} - \epsilon_{\nu,N_{\mathrm{eff}} = 3.044})}$.

Unsurprisingly, the effect of altering $G_{\mathrm{eff}}$ for fixed $N_{\mathrm{eff}}$ on the phase $\Delta \ell$ (or $\Delta k$) is similar to the effect seen in \cite{choi2018probing} on the phase shift when free-streaming is delayed. For neutrinos with more greatly delayed free-streaming (for larger $G_{\mathrm{eff}}$), there is a decreasing amplitude for the phase shift and a turnover scale emerges which may be related to the scale identified in \cite{choi2018probing} as the horizon corresponding to the delayed neutrino free-streaming.

\section{Can we constrain $G_{\mathrm{eff}}$ from the phase of the BAO?}\label{sec::constraintsBAO}

Having studied the impact of neutrino self-interactions on the phase shift, we now examine whether its detection can be used to constrain non-standard neutrino interactions. Unlike constraints on $G_{\mathrm{eff}}$ that rely on the direct study of cosmological observables, such as the CMB and matter power spectra, analyzing the distinct imprint of phase shift has the advantage of being less degenerate with other cosmological parameters, e.g., $A_s$ and $n_s$. However, we \textit{can} expect it to be highly degenerate with the impact of $N_{\mathrm{eff}}$ on the amplitude of the phase, or any other parameter that alters $r_s$. Despite this limitation, the non-linear dependence of $G_{\mathrm{eff}}$ on $k$ may still provide additional constraining power. 

A typical BAO analysis involves comparing the BAO oscillations $\mathcal{O}(k)$ to a template for $\mathcal{O}(k)$ that is computed at some fixed `template' cosmology, and then finding the best fit to the `distortion' parameters $\alpha_{\parallel}$ and $\alpha_{\perp}$, which shift the positions of the BAO oscillations in the template to match them to the data. These parameters ultimately capture two impacts; the first is the Alcock-Paczynski (AP) effect \citep{alcock1979evolution}, which leads to the observed distances to galaxies (and thus the BAO feature) to differ from the true distances when one assumes a fiducial cosmology to transform galaxy redshifts into physical distances. The second is the impact of the choice of cosmology in the BAO template, which leads the physical scale $r_s$ to differ from that of the true cosmology. As such, $\alpha_{\parallel}$ and $\alpha_{\perp}$ allow one to determine how much the fiducial cosmology and template cosmology differ from the true cosmology, and constrain parameters that alter the expansion rate.\footnote{These parameters enter by shifting $\mathcal{O}(k) \rightarrow \mathcal{O}(k')$; see more details in ref.~\cite{whitford2024constraining} for the equations relating the BAO distortions parameters to $k$, $k'$, and for the equations that define $\alpha_{\parallel}$, $\alpha_{\perp}$.} The growth rate of structure multiplied by $\sigma_8$, $f\sigma_8$ and the galaxy bias $b$ also enter as free parameters in the analysis; these parameters change the impact of redshift space distortions through the Kaiser effect \citep{kaiser1987clustering}. Including the parameter $\beta_{\phi}$ extends upon a typical BAO analysis, but allows one to capture the impact of free-streaming neutrinos on the BAO phase (see more details in \cite{baumann2019first, whitford2024constraining}). Here, we go a step further by including the impact of $G_{\mathrm{eff}}$ on the BAO phase via the parameters $A(G_{\mathrm{eff}})$ and $B(G_{\mathrm{eff}})$. 

\subsection{Profile of $\chi^2$}\label{subsec:profilechi2BAO}

\begin{figure}[h!]
    \begin{subfigure}{0.5\textwidth} \includegraphics[width=0.97\linewidth]{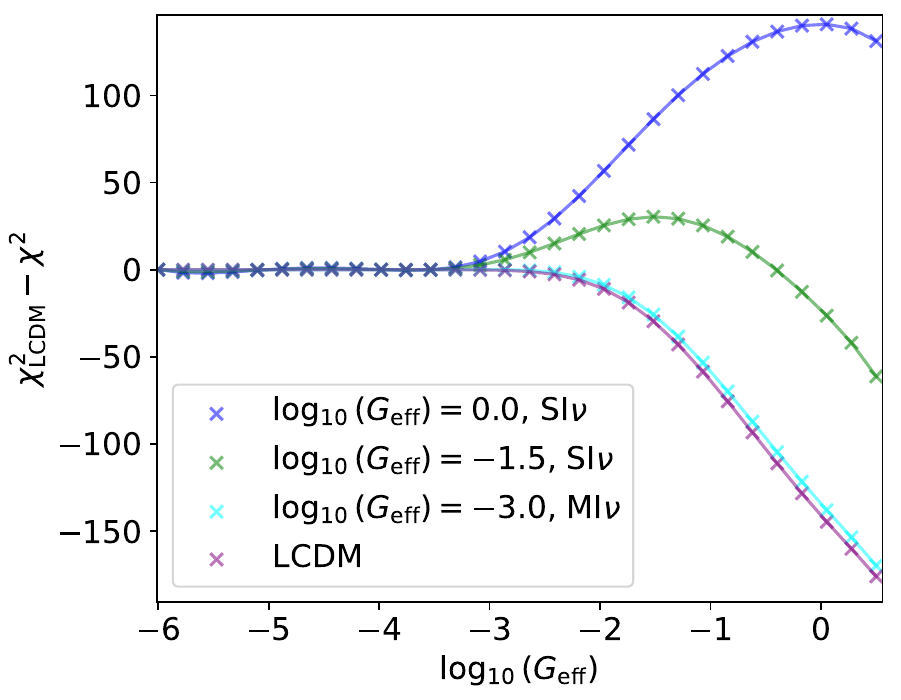}
    \caption{}
    \label{fig:chisquare_profile_BAO}
    \end{subfigure}\begin{subfigure}{0.5\textwidth} \includegraphics[width=1.04\linewidth]{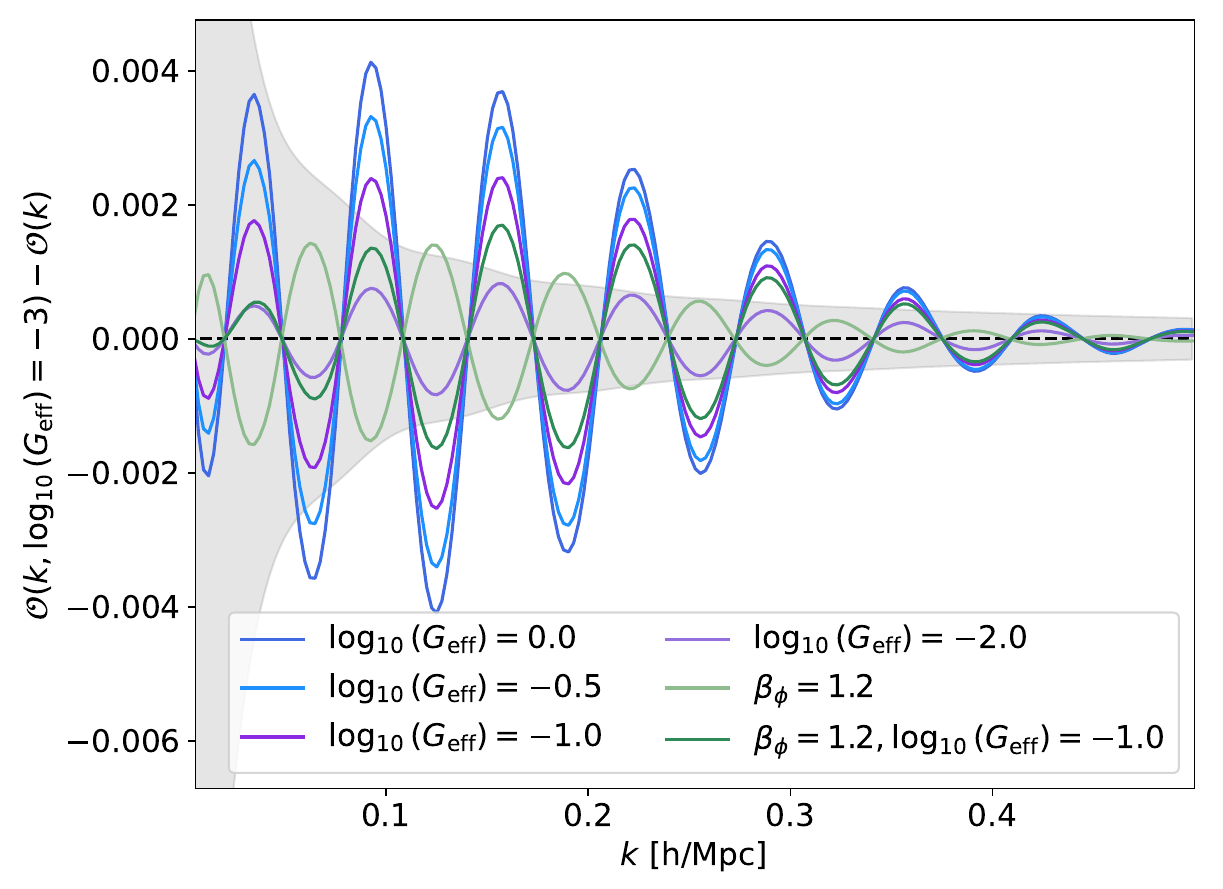}\caption{}\label{fig:errorbarsBAO}
    \end{subfigure}\caption{a) The difference between the $\chi^2$ goodness-of-fit with $\Lambda$CDM and the $\chi^2$ goodness-of-fit for a model with neutrino self-interactions parameterized by $G_{\mathrm{eff}}$, as a function of $G_{\mathrm{eff}}$. We show two models for strong (SI$\nu$) interactions, and one for moderate interactions (MI$\nu$). This shows the profile for BAO data with a fiducial model that is indicated by the legend. It is the best case result, for a volume-limited survey. b) A plot of the BAO signal for a cosmology in which $\log_{10}{(G_{\mathrm{eff}})} = -3$, after subtracting the BAO signal from cosmologies with varying phase shifts due to variations in $G_{\mathrm{eff}}$ or $\beta_{\phi}$. The grey shaded region shows the theoretical error bars for a volume limited survey, in which the volume covers the range $z = 0 - 2$, and the bins are of width $\Delta k = 0.0025\, h\, \mathrm{Mpc}^{-1}$. }
    \centering
\end{figure}

We start with considering the BAO oscillations measured from a cosmic-variance (or volume) limited survey, in order to determine if any information is available in the best-case scenario. We first consider what ability there is to constrain $G_{\mathrm{eff}}$ alone from the BAO phase, which is illustrated by Figure~\ref{fig:chisquare_profile_BAO}. This illustrates the constraining power for a volume-limited survey that is able to measure the BAO feature in redshift bins of width $\Delta z = 0.1$ spanning $z = 0 - 2$, with a sky area of $\sim 30,000$ square degrees. We consider the constraining power in a survey with $k_{\mathrm{min}}, k_{\mathrm{max}} = 0.001h, 0.5h$ Mpc$^{-1}$ with bins of width $\Delta k = 0.0025h$ Mpc$^{-1}$. In this first simple analysis, we fix $G_{\mathrm{eff}}$ and allow the usual BAO parameters in each redshift bin to vary ($f\sigma_8$, $b\sigma_8$, $\alpha_{\parallel}$, $\alpha_{\perp}$). In section~\ref{subsec:forecastsBAO}, we obtain an alternative calculation of the significance of a measurement compared to $\Lambda$CDM. For BAO data with a true cosmology corresponding to some choice of $G_{\mathrm{eff}}$, we effectively plot the `profile likelihood'---the difference between the $\chi^2$ goodness-of-fit of the data assuming a $\Lambda$CDM model compared to models with different $G_{\mathrm{eff}}$. The $\Delta \chi^2$ is essentially the difference between a null hypothesis as $\Lambda$CDM and a model with varying $G_{\mathrm{eff}}$, $\Delta \chi^2 = \chi^2_{\mathrm{LCDM}} - \chi^2$, allowing us to determine a confidence interval for a measurement. In order to compute each $\Delta \chi^2$, we simply find the optimal \{$f\sigma_8$, $b\sigma_8$, $\alpha_{\parallel}$, $\alpha_{\perp}$\}, in each redshift bin for our `data' when one assumes $\Lambda$CDM or various models with varying $G_{\mathrm{eff}}$. The minimum $\chi^2$ is obtained with a simple optimization with respect to the `data', using \textsc{scipy.optimize.minimize}. We model the variance of the power spectrum at each $z$ using the result of \citep{tegmark1997measuring} as $\frac{2}{N_{\mathrm{modes}}}(P(k,\mu,z))^2$, where $N_{\mathrm{modes}}$ is the number of modes that can be observed in a particular volume, which depends on $\bar{n}$, the number density of galaxies. For a volume-limited survey, the terms involving the number densities vanish. In this analysis, we expect to see a peak at the model $G_{\mathrm{eff}}$ that matches the truth most closely. A sharper peak indicates better ability to constrain $G_{\mathrm{eff}}$. 
% For our calculations we use the template 
This analysis use the template Equation~\eqref{eq:templatelog10geff_BAO} to model the impact of altering $G_{\mathrm{eff}}$ on the BAO signal, explicitly ensuring that only information available in the phase shift about $G_{\mathrm{eff}}$ is included in the calculation. This is also demonstrated by Figure~\ref{fig:errorbarsBAO} which shows the difference between the BAO spectrum when the phase is altered due to variations in $G_{\mathrm{eff}}$ (and for comparison $\beta_{\phi}$), with the shaded region showing error bars on the BAO power spectrum. 

% By eye, 
As shown in Figure~\ref{fig:chisquare_profile_BAO}, our results suggest that strong interactions (SI$\nu$, $\log_{10}{(G_{\mathrm{eff}})} > -2$) are able to be constrained by the cosmic-variance limited BAO data---we will explore what significance these results correspond here and also in the next section. In such cases, the $\Delta \chi^2$ peaks about the value of $G_{\mathrm{eff}}$ that matches the fiducial model employed to generate `data'. Moderate interactions (MI$\nu$, $\log_{10}(G_{\mathrm{eff}}) \leq -2$) are difficult to distinguish from the case of no neutrino self-interactions, given the nearly flat slope of $\Delta \chi^2$. Nevertheless, a preference against strong interactions remains possible, due to the sharp drop in $\Delta \chi^2$ at larger $G_{\mathrm{eff}}$. These results align with expectations: the phase-shift signal is more pronounced for strong interactions, while very weak interactions have little impact. 

To assess the statistical significance of our results, we compute the probability associated with $\Delta \chi^2$ under the null hypothesis. Assuming that $\Delta \chi^2$ follows a $\chi^2$ distribution with 80 degrees of freedom ($20$ redshift bins $\times 4$ parameters per bin), the probability $p$ of obtaining a larger $\Delta \chi^2$ value under the null hypothesis ($\Lambda$CDM) is $p = 1 - \mathrm{CDF}_{k=1}(\Delta \chi^2)$, where CDF denotes the cumulative distribution function of the $\chi^2$ distribution. The corresponding significance in units of standard deviations, $\sigma$, can then be obtained by applying the inverse normal CDF at $1 - p$ (for a one-tailed test).

We find that a cosmological scenario with strongly interacting neutrinos (dark blue line in Figure~\ref{fig:chisquare_profile_BAO}) following $\log_{10}{(G_{\mathrm{eff}})} = 0.0$ can be detected via phase shift analysis with a significance of $\sim 4\sigma$. However, if the self-interaction strength of the data is slightly weaker with $\log_{10}{(G_{\mathrm{eff}})} = -1.5$ (green line), the $\Lambda$CDM and self-interacting neutrino models become essentially indistinguishable. The same is true for moderate interacting scenarios with ($\log_{10}{(G_{\mathrm{eff}})} = -3$, cyan line). Nonetheless, in the case of a Universe with $\log_{10}{(G_{\mathrm{eff}})} = -3$ or standard model neutrinos, an analysis of the phase shift allows to rule out much stronger interactions, $\log_{10}{(G_{\mathrm{eff}})} = 0.0$, at a significance also of $\sim 3.8\sigma$. Overall, this suggests that we cannot strongly constrain any interactions with the BAO phase shift alone unless they are very strong interactions, or we are trying to show that strong interactions are not preferred by the data. In this case, the BAO signal can be useful. However, we may at least be able to obtain information from the BAO that could complement CMB constraints, which we expect to have much better constraining power; this will be explored in section~\ref{sec:constraintsCMB}. We will also explore how our ignorance about $\beta_{\phi}$ will impact these constraints in the next section.

\subsection{Fisher matrix forecasts}\label{subsec:forecastsBAO}

To understand the constraining power of the BAO phase shift when $G_{\mathrm{eff}}$ (and $\beta_{\phi}$) are free parameters, we study Fisher matrix forecasts for a volume-limited BAO survey. For a galaxy survey our observable is the Fourier-space galaxy density field $\delta(k)$ which has a mean $\mu = \bar{\delta}(k) = 0$. We can model the covariance $\Sigma$ for this as $\Sigma = P(k,z,\mu)$, where the shot noise is $\sim 0$ for a volume-limited survey.\footnote{$\Sigma$ and $\mu$ take a matrix form and vector form when computing the Fisher information for multiple tracers.} $P(k)$ is the matter power spectrum in redshift space. The Fisher information for our observable, which is a multivariate Gaussian distribution of our parameter vector $\mathbf{x}_i$ (containing our cosmological parameters of interest) can be written as 
\begin{align}
    F_{ij} & = \frac{d\mu}{d\theta_i} \Sigma^{-1} \frac{d\mu}{d\theta_j} + \frac{1}{2}\mathrm{tr} \left( \Sigma^{-1} \frac{d\Sigma}{d\theta_i} 
 \Sigma^{-1} \frac{d\Sigma}{d\theta_j}  \right) \\ \nonumber 
 & = \frac{1}{2}\mathrm{tr} \left( \Sigma^{-1} \frac{d\Sigma}{d\theta_i} 
 \Sigma^{-1} \frac{d\Sigma}{d\theta_j}\right). 
\end{align}
The first term is zero for $\mu = \bar{\delta}$. Fisher information is additive, and thus one can integrate the information over all modes $k$, redshifts $z$ and angles $\mu$ observed in the survey to obtain the information over a 3D volume. Thus this equation becomes a triple integral to capture the information over all angles and accessible scales,
\begin{equation}
    F_{ij} = \int_0^{r_{\mathrm{max}}(z)} d^3r \int_{k_{\mathrm{min}}}^{k_\mathrm{max}} d^3k \int_0^1 d\mu \frac{1}{2}\mathrm{tr} \left( \Sigma^{-1} \frac{d\Sigma}{d\theta_i} 
 \Sigma^{-1} \frac{d\Sigma}{d\theta_j}  \right).
\end{equation}

To compute the Fisher information, we modified the forecasting code \textsc{GoFish} \citep[which produces the power spectrum with \textsc{CAMB},][]{lewis2011camb} to include the phase shift amplitude for standard model neutrinos, $\beta_{\phi}$, and $\log_{10}{(G_{\mathrm{eff}})}$.\footnote{The modified code can be found at \url{https://github.com/abbew25/GoFish}.} The original code was developed to produce forecasts for the DESI survey, so more details about the modelling in redshift-space can be found in \cite{adame2024validation}. The Cramér-Rao bound states that the minimal (statistical) variance of a parameter given an observable is bound by the inverse of the Fisher information. This allows one to determine a forecast for a parameter of interest from a galaxy survey with a given signal-to-noise ratio. 

We assume a volume-limited survey as previously described in section~\ref{subsec:profilechi2BAO}, with a sky area of 30,000 square degrees, with 20 redshift bins from $z = 0$ to $z = 2$. We allow $k_{\mathrm{min}}, k_{\mathrm{max}} = 0.001h, 0.5h$ Mpc$^{-1}$. For our Fisher forecasts we set the galaxy bias $b_g = 1$ at $z = 0$ and allow it to vary proportionally to $\frac{\sigma_8(0)}{\sigma_8(z)}$ with $z$. For all forecasts, we allow the following to be free parameters: $b_g(z)_i \sigma_8(z)_i$ for each redshift bin $i$, $f \sigma_8(z)_i$ where $f$ is the growth rate of structure, and $\alpha_{\parallel}$ and $\alpha_{\perp}$ (the anisotropic BAO distortion parameters). We compute the constraining power on $\log_{10}{(G_{\mathrm{eff}})}$, for different choices of $\log_{10}{(G_{\mathrm{eff}})}$. We also consider the case where $\beta_{\phi}$ (which changes due to $N_{\mathrm{eff}}$ and increases the amplitude of the phase shift signal) is allowed to vary freely. Results are shown in Table~\ref{tab:BAOvolumelimitedresultsz0_2}. For a more realistic scenario, we also consider the ability of DESI to constrain neutrino self-interactions from the BAO signal alone in the full five-year data, which is shown in Table~\ref{tab:BAODESI_validationnumbers}. For very large error bars ($\geq 15)$ we simply state that $G_{\mathrm{eff}}$ is not constrained in these tables. For our DESI forecasts, we use a sky area of 14,000 square degrees and set $k_{\mathrm{min}}, k_{\mathrm{max}} = 0.001h, 0.3h$ Mpc$^{-1}$. For the number densities of galaxies in BGS, ELGs, LRGs and QSO with $z$ and their bias, we follow the values provided in \cite{adame2024validation}. For multitracer galaxy surveys, the Fisher Information can be computed by simply extending $\mu$ to a vector $\boldsymbol{{\mu}} = [\delta_0, \delta_1 ...,\delta_N]$ for $N$ tracers. $\Sigma$ is extended to a $N \times N$ covariance matrix $\boldsymbol{\Sigma}_{ij}$ in which the diagonal terms give the variance on each tracer's density field and the covariance between different tracers is given by the off-diagonals. We treat the constraints on $b_g(z)_i \sigma_8(z)_i$, $f \sigma_8(z)_i$, $\alpha_{\parallel}(z)$ and $\alpha_{\perp}(z)$ independently for each tracer in each redshift bin, which allows for stronger constraints by cross-correlating information between different tracers. 

\begin{table}[h!]
    \caption{Fisher forecasts for the ability to constrain $G_{\mathrm{eff}}$, for a few different models, for a volume limited survey. Constraints are shown when one fixes $\beta_{\phi} = 1$ (for $N_{\mathrm{eff}} = 3.044$) or when it is allowed to vary (effectively allowing $N_{\mathrm{eff}}$ to change the BAO phase amplitude freely). In brackets, we show the number of $\sigma$ each constraint gives for $\log_{10}(G_{\mathrm{eff}})$ relative to $\log_{10}(G_{\mathrm{eff}}) = -3$ (we choose this value since it cannot realistically be distinguished from smaller choices of $G_{\mathrm{eff}}$). }
    \centering
    \begin{tabular}{|c|c|c|} \hline 
        Model & $\sigma(\log_{10}{(G_{\mathrm{eff}})})|_{\beta_{\phi}=1}$ & $\sigma(\log_{10}{(G_{\mathrm{eff}})})_{{\beta_{\phi}} \mathrm{free}}$  \\ \hline 
        SI$\nu$, $\log_{10}{(G_{\mathrm{eff}})} = 0.25$ & $\pm 0.86$ (3.96) & $\pm 6.59$ (0.0) \\
        SI$\nu$, $\log_{10}{(G_{\mathrm{eff}})} = 0$ & $\pm 0.86$ (3.49) & $\pm 13.47$ (0.0) \\
        SI$\nu$, $\log_{10}{(G_{\mathrm{eff}})} = -1.5$ & $\pm 1.62$ (0.93) & $\pm 8.82$ (0.0) \\
        MI$\nu$, $\log_{10}{(G_{\mathrm{eff}})} = -3$ & unconstrained & unconstrained \\  \hline 
    \end{tabular}
    \label{tab:BAOvolumelimitedresultsz0_2}
\end{table}

\begin{table}[h!]
    \caption{Fisher forecasts for the ability to constrain $G_{\mathrm{eff}}$, for a few different models, for the DESI 5-year dataset. The columns are the same as those in Table~\ref{tab:BAOvolumelimitedresultsz0_2}. }
    \centering
    \begin{tabular}{|c|c|c|} \hline  
        Model & $\sigma(\log_{10}{(G_{\mathrm{eff}})})|_{\beta_{\phi}=1}$ & $\sigma(\log_{10}{(G_{\mathrm{eff}})})_{{\beta_{\phi}} \mathrm{free}}$  \\ \hline 
        SI$\nu$, $\log_{10}{(G_{\mathrm{eff}})} = 0.25$ &  $\pm 2.07$ (1.57) & $\pm 13.76 $ (0.0) \\
        SI$\nu$, $\log_{10}{(G_{\mathrm{eff}})} = 0$ & $\pm 2.19 $ (1.37) & unconstrained\\
        SI$\nu$, $\log_{10}{(G_{\mathrm{eff}})} = -1.5$ & $\pm 4.24$ (0.0) & $\pm 12.3$ (0.0)  \\  \hline 
    \end{tabular}
    \label{tab:BAODESI_validationnumbers}
\end{table}

The results in Table~\ref{tab:BAOvolumelimitedresultsz0_2} are consistent with our results from the $\chi^2$ profile analysis; it is possible to obtain constraints on $G_{\mathrm{eff}}$, but only for very strong interactions, and in the case one can assume that $N_{\mathrm{eff}}$ is known well enough to fix $\beta_{\phi} = 1$. Interestingly, our detection significance of $\sim 3.5\sigma$ for the case of a model with strong interactions ($\log_{10}{(G_{\mathrm{eff}})} = 0$) is similar to what we obtained from the $\chi^2$ profile analysis, which found a detection significance of $\sim 4\sigma$. For weaker interactions, or when we allow $\beta_{\phi}(N_{\mathrm{eff}})$ to vary, the constraining power decreases. This is largely because the derivative $\frac{d\mathcal{O}(k)}{dG_{\mathrm{eff}}}$ is extremely small since the phase shift does not change much for weak self-interactions (see Figure~\ref{fig:templatefitlog10Geff_BAO}; the Fisher information directly depends on this), and due to the fact that the amplitude modulations by $A(G_{\mathrm{eff}})$ is degenerate with $\beta_{\phi}$. Moreover, the modulation of the phase-shift amplitude by $G_{\mathrm{eff}}$ is less significant than the effect of varying $N_{\mathrm{eff}}$ and thus $\beta_{\phi}$. However, larger values of $N_{\mathrm{eff}}$ imply larger $\beta_{\phi}$ and thus a greater amplitude of the phase shift signal, in which case the amplitude modulation by $G_{\mathrm{eff}}$ may become more important, restoring some constraining power on $G_{\mathrm{eff}}$. 

While these results suggest that we are able to gain little information about $G_{\mathrm{eff}}$ at all if only weak interactions occur, our results from Figure~\ref{fig:chisquare_profile_BAO} suggest there is sufficient information to conduct a model comparison with real data, providing a potential avenue to rule out strong interactions (or more extreme scenarios). It is possible to estimate the significance level at which stronger interactions can be disfavoured using the Fisher information $\boldsymbol{F}$; one can approximately compute a $\Delta \chi^2$ between two models as $\Delta \chi^2 \approx \Delta \boldsymbol{\theta}^T \boldsymbol{F} \Delta\boldsymbol{\theta}$. If we take the Fisher information from our constraints for a model with $\log_{10}{(G_{\mathrm{eff}})} = -3$, and compare to a model with $\log_{10}{(G_{\mathrm{eff}})} = 0$, this gives $\Delta \chi^2 \approx 6.75$, corresponding to a preference of the weaker interaction model over the other at the level of $\sim 0.8\sigma$ (here the number of degrees of freedom is just one for our single parameter that changes between the two models, $G_{\mathrm{eff}}$, leaving all other parameters fixed). Compared to our profile analysis, this is smaller (we found earlier a significance of $\sim 3.8\sigma$ compared to $\Lambda$CDM), although this calculation does not take into account variation of other parameters.

The results of table~\ref{tab:BAODESI_validationnumbers}, that forecast for upcoming data from DESI, find the constraints are (as we expect) weaker than those shown for a volume limited survey, but there is still some ability to constrain $G_{\mathrm{eff}}$ from the phase shift for very strong interactions if we assume a fixed value of $N_{\mathrm{eff}}$; otherwise, as for the volume limited case, the constraints are not useful from BAO data alone. In Section~\ref{comboconstraints}, we will explore whether there are improved constraints from upcoming BAO data from a survey such as DESI, if it is combined with the constraining power of a CMB experiment.

\section{Can we constrain $G_{\mathrm{eff}}$ from the phase of the CMB power spectra?}\label{sec:constraintsCMB}

\subsection{Profile of $\chi^2$}

As for the BAO, we conduct a calculation of an effective `profile likelihood' of the $\chi^2$ goodness-of-fit to the data, by computing $\Delta \chi^2$ between the model of the null hypothesis ($\Lambda$CDM) and models with varying $G_{\mathrm{eff}}$. As before for the BAO, the $\Delta \chi^2$ involves varying a number of parameters that are standard for a CMB analysis while leaving $G_{\mathrm{eff}}$ fixed. %We vary $G_{\mathrm{eff}}$ for various `datasets' corresponding to some true value of $G_{\mathrm{eff}}$, and expect to see a peak about this $G_{\mathrm{eff}}$ that corresponds to that of the data. 
We vary $G_{\mathrm{eff}}$ across simulated datasets generated with different true values of $G_{\mathrm{eff}}$. In each case, we expect the likelihood to peak near the input value. A sharper peak indicates a stronger detection. For simplicity, in this section we just consider the information in the CMB temperature-temperature (TT) power spectrum, and consider a cosmic variance limited survey (instrument noise is neglected) with the same sky area of $\sim 30000$ square degrees. We consider $\ell$ in the range $\ell = [2, 2500]$, and use the raw CMB power spectra. As before, a Fisher matrix analysis of the ability to constrain will also be given in the next section, which will include information from the polarization and temperature-polarization cross power spectra.

\begin{figure}[h!]
    \begin{subfigure}{0.48\textwidth} \includegraphics[width=1\linewidth]{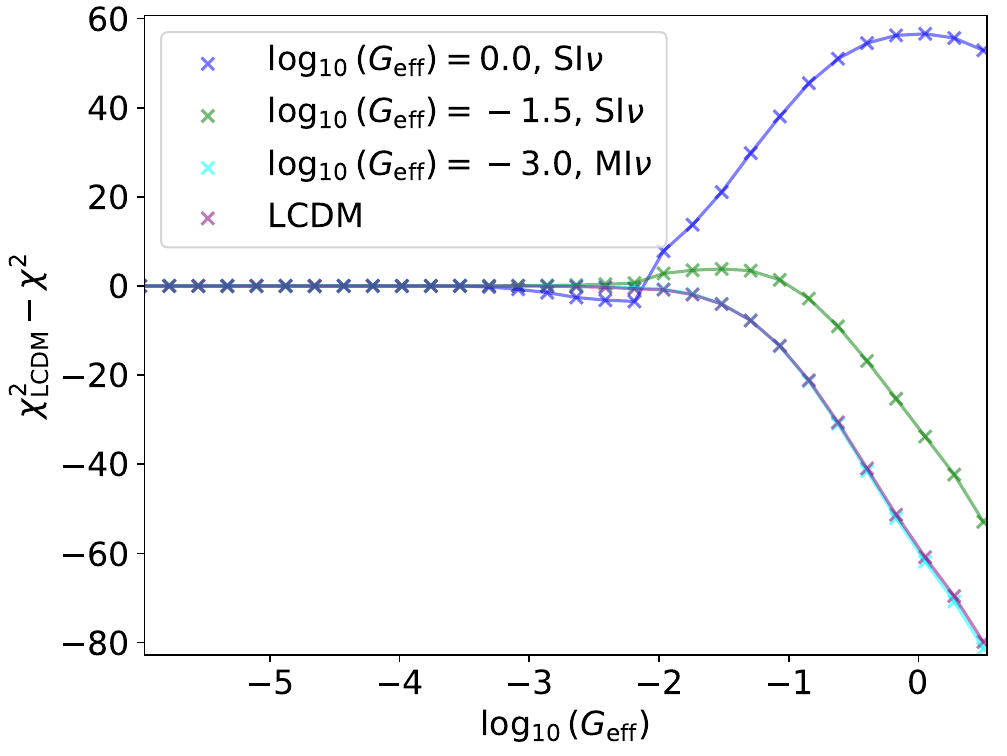}
    \caption{}
    \label{fig:CMB_profilechisquared}
    \end{subfigure}\begin{subfigure}{0.52\textwidth} \includegraphics[width=1\linewidth]{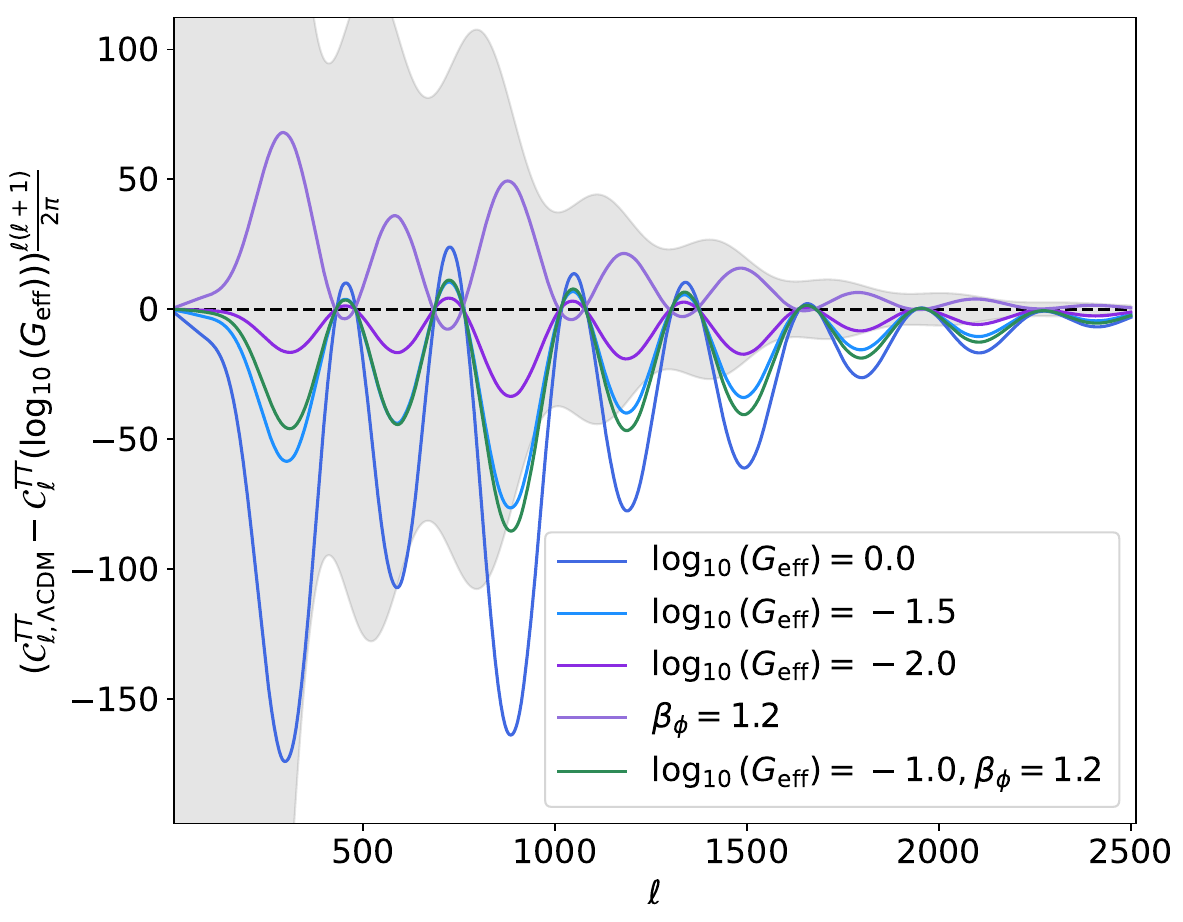}
    \caption{}
    \label{fig:CMB_CV_errorbars}
    \end{subfigure}
    \caption{{a) The difference between the $\chi^2$ goodness-of-fit with $\Lambda$CDM and the $\chi^2$ goodness-of-fit for a model with neutrino self-interactions parameterized by $G_{\mathrm{eff}}$. As in Figure~\ref{fig:chisquare_profile_BAO}, this shows the profile (but with CMB data) for a fiducial model that is indicated by the legend. This is the best case result for a cosmic-variance limited survey. The constraints for data corresponding to $\log_{10}(G_{\mathrm{eff}}) = -3$ and to $\Lambda$CDM are statistically indistinguishable, so the cyan and pink lines overlap. b) A plot of the CMB TT spectrum in a cosmology for which $\log_{10}{(G_{\mathrm{eff}})} = -3$, after subtracting the CMB signal from cosmologies with varying phase shifts due to variation in $G_{\mathrm{eff}}$ or $\beta_{\phi}$. The grey shaded region shows the theoretical error bars for a cosmic-variance limited CMB experiment, in which the sky area covers $\sim 30,000$ square degrees.}}
\end{figure}

To calculate the $\chi^2$ we compare the `observed' $C_{\ell}$s to a model; we assume the variance on $C_{\ell}$ as $\frac{2}{(2\ell + 1)f}C_{\ell}^2$, where $f$ is the fraction of sky coverage. The minimum $\chi^2$ is found with respect to the `data' also using \textsc{scipy.optimize.minimize}. For our analysis, we allow the following cosmological parameters to vary: $\{ h, \omega_{\mathrm{b}} = \Omega_{\mathrm{b}}h^2, \omega_{\mathrm{c}} = \Omega_{\mathrm{c}}h^2, \ln{(10^{10} A_s)}, n_s, \tau \}$. Since generating the power spectrum is slow in our minimization algorithm, we use the \textsc{cosmopower} emulator for the CMB TT power spectrum \citep{spuriomancini2021} (and as for the BAO, use our template to include the impact of $G_{\mathrm{eff}}$ on the phase). Figure~\ref{fig:CMB_profilechisquared} shows the profile likelihood, and Figure~\ref{fig:CMB_CV_errorbars} shows a comparison of how the phase shift changes between different models with varying $G_{\mathrm{eff}}$ and $\beta_{\phi}$. The grey shaded region shows the uncertainty on a CV limited CMB experiment TT power spectrum. 

From Figure~\ref{fig:CMB_profilechisquared}, we can see that the $\Delta \chi^2$ indicates an ability to distinguish $\Lambda$CDM and models with neutrino self-interactions in the CMB, with similar trends to that seen for the BAO. The $\Delta \chi^2$ reaches a maximum for the most strongly interacting model (shown as the dark blue line in Figure for the case the `data' corresponds to a model with $\log_{10}{(G_{\mathrm{eff}})} = 0.0$). The green line that corresponds to a model with $\log_{10}{(G_{\mathrm{eff}})} = -1.5$ shows the expected peak (albeit a small peak) about -1.5. For models with weaker interactions, there is poorer ability to differentiate from $\Lambda$CDM, but there is some ability to distinguish these models from stronger interactions. Following the same procedure as before to interpret the $\Delta \chi^2$ as a detection significance (now from a $\chi^2$ distribution with $k=6$ degrees of freedom), we find for the case the data has a cosmology corresponding to $\log_{10}{(G_{\mathrm{eff}})} = 0.0$, the detection significance is $\sim 6.2\sigma$ with respect to $\Lambda$CDM. Similar to the case shown for the BAO, the model with $\log_{10}{(G_{\mathrm{eff}})} = -1.5$ or smaller values cannot be distinguished from $\Lambda$CDM based on this analysis. However, taking the model in which $\log_{10}{(G_{\mathrm{eff}})} = -3$, we can show that $\log_{10}{(G_{\mathrm{eff}})} = -3$ is preferred over stronger interactions, corresponding to $\log_{10}{(G_{\mathrm{eff}})} = 0.0$ at a significance of $\sim 6.6\sigma$. Overall, this result is similar to what we find in the BAO profile analysis, except that the detection significance is greater for very strong interactions. 

\subsection{Fisher matrix forecasts}

In the Fisher matrix analysis, we include not only the TT spectrum, but also the EE and TE spectra to determine the full constraining power of the CMB in different cases. We also allow $G_{\mathrm{eff}}$ to be a free parameter (and in separate cases $\beta_{\phi}$). For a CMB experiment, our observable is the Fourier-space temperature fluctuation field across different angles, $\frac{\Delta T(\theta)}{\bar{T}}$; the mean vanishes, like in the case for the matter field $\delta$. However, here we use the covariance matrix of the CMB power spectrum rather than the map itself. This depends on the CMB auto-power spectrum of temperature fluctuations, $C^{TT}_{\ell}$, giving
\begin{equation}
    \Sigma_{TT} = \frac{2}{(2\ell + 1) f} (C^{TT}_{\ell} + N^{TT}_{\ell})^2. 
\end{equation}
$N^{TT}_{\ell}$ is the noise on the CMB power spectrum that comes from the instruments. Obtaining the full Fisher information involves summing over all the accessible angular modes $\ell$ and power spectra $\{a,b\}$ as 
\begin{equation}
    F_{ij} = \sum_{a,b} \sum_{\ell_{\mathrm{min}}}^{\ell_{\mathrm{max}}} \left( \frac{dC_{\ell}^a}{d\theta_i} 
 \boldsymbol{\Sigma}^{-1}_{a,b} \frac{dC_{\ell}^b}{d\theta_j}  \right). 
\end{equation}
Similar to the case of computing the Fisher information for multiple galaxy survey tracers, the analysis can be extended to include information from the polarization power spectrum and temperature-polarization cross-power spectrum, thus we have to consider the power spectra for $\{ TT, TE, EE\}$. In this case the covariance matrix is given by
\begin{align}
    \boldsymbol{\Sigma} &= \frac{2}{(2\ell + 1)f_{\mathrm{sky}}} \\ \nonumber
    & \times \begin{bmatrix} (C^{TT}_{\ell} + N^{TT}_{\ell})^2 & C^{TE}_{\ell}C^{TE}_{\ell} & (C^{TT}_{\ell} + N^{TT}_{\ell})C^{TE}_{\ell} \\ 
    C^{TE}_{\ell}C^{TE}_{\ell} & (C^{EE}_{\ell} + N^{EE}_{\ell})^2 & (C^{EE}_{\ell} + N^{EE}_{\ell})C^{TE}_{\ell} \\ 
    (C^{TT}_{\ell} + N^{TT}_{\ell})C^{TE}_{\ell} & (C^{EE}_{\ell} + N^{EE}_{\ell})C^{TE}_{\ell} & \frac{1}{2}(C^{TE}_{\ell}C^{TE}_{\ell} + (C^{TT}_{\ell} + N^{TT}_{\ell})(C^{EE}_{\ell} + N^{EE}_{\ell}))
    \end{bmatrix}. 
\end{align} We assume instrument noise is not correlated between the spectra, so $N^{TE}_{\ell} = 0$. As in the previous section, we produce forecasts for a survey with a sky area of $\sim$ 30,000 square degrees, which is cosmic-variance limited, giving $N^{\mathrm{TT}}_{\ell} = N^{\mathrm{EE}}_{\ell} = 0$. We use the range $\ell = [2, 2500]$. Aside from $G_{\mathrm{eff}}$ we include the following free-parameters in the forecast; $100\theta_s$, $100\omega_{\mathrm{b}} = 100\Omega_{\mathrm{b}}h^2$, $\omega_{\mathrm{c}} = \Omega_{\mathrm{c}}h^2$, $\ln{(10^{10}A_s)}$, $n_s$, $\tau$. We also include $\beta_{\phi}(N_{\mathrm{eff}})$ in separate forecasts. As for the BAO, we explicitly include information about $G_{\mathrm{eff}}$ from the phase shift only by using the template to change the CMB phase shift due to variations in $G_{\mathrm{eff}}$ and $\beta_{\phi}$ when these parameters are varied. In order to produce the forecasts, we modified the \textsc{GoFish} code to produce forecasts for CMB data, which is able to use CAMB or CLASS.\footnote{\url{https://github.com/abbew25/CMB-fish}} For our forecasts, we neglect CMB lensing and use the raw CMB power spectra.   

In Table~\ref{tab:CMB_forecasts_CV}, we show forecasts for a cosmic variance limited experiment. We are able to obtain forecasts for more weakly interacting models; compared to the case for the BAO the CMB is much more constraining. As is the case for the BAO, it is easier to constrain the interactions from the phase shift when they are stronger, as $\frac{dC(\ell)}{dG_{\mathrm{eff}}}$ is larger. 

\begin{table}[h!]
    \caption{Fisher forecasts for the ability to constrain $G_{\mathrm{eff}}$, for a few different models, for a cosmic variance limited CMB measurement of the TT, EE and TE power spectra. The columns are the same as those in Table~\ref{tab:BAOvolumelimitedresultsz0_2}. The detection significance is given in $\sigma$ relative to a model with $\log_{10}{(G_{\mathrm{eff}})} = -3.5$, in which the phase shift is essentially indistinguishable from $\Lambda$CDM. }
    \centering
    \begin{tabular}{|c|c|c|} \hline 
        Model & $\sigma(\log_{10}{(G_{\mathrm{eff}})})|_{\beta_{\phi}=1}$ & $\sigma(\log_{10}{(G_{\mathrm{eff}})})_{{\beta_{\phi}} \mathrm{free}}$  \\ \hline 
        %SI$\nu$, $\log_{10}{(G_{\mathrm{eff}})} = 0.25$ &  \\
        SI$\nu$, $\log_{10}{(G_{\mathrm{eff}})} = 0$ & $\pm 0.02 (>5)$ & $\pm0.06 (>5)$   \\
        SI$\nu$, $\log_{10}{(G_{\mathrm{eff}})} = -1.5$ & $\pm 0.04 (>5)$ & $\pm0.11 (>5)$ \\ 
        MI$\nu$, $\log_{10}{(G_{\mathrm{eff}})} = -3$ & $\pm 0.27 (1.85) $ & $\pm0.49 (1.02)$ \\ \hline 
    \end{tabular}
    \label{tab:CMB_forecasts_CV}
\end{table}

These results suggest that upcoming CMB experiments will be able to use the phase shift to constrain $G_{\mathrm{eff}}$, even if the constraints will be weaker given realistic noise properties for upcoming surveys; for a cosmic-variance limited survey, the CMB will be able to achieve extremely strong constraints ($> 5\sigma$) on neutrino self-interactions if they exist. For very weak interactions or a fit to cosmology that is consistent with $\Lambda$CDM, it should be possible to disfavour stronger interactions using just the CMB. In Table~\ref{tab:CMB_forecasts_Planck} we show forecasts for a \textit{Planck}-like experiment, using the noise properties assumed in \cite{font2014desi}. We assume $f = 0.7$, $\ell = [2, 2500]$ for the TT spectrum and $[2, 2000]$ for the EE and TE spectrum, and we use the unlensed spectra. 

\begin{table}[h!]
    \caption{Fisher forecasts for the ability to constrain $G_{\mathrm{eff}}$, for a few different models, for a \textit{Planck}-like CMB measurement of the TT, EE and TE power spectra. The columns are the same as those in Table~\ref{tab:BAOvolumelimitedresultsz0_2}. The detection significance is given in $\sigma$ relative to a model with $\log_{10}{(G_{\mathrm{eff}})} = -3.5$, in which the phase shift is essentially indistinguishable from $\Lambda$CDM.  }
    \centering
    \begin{tabular}{|c|c|c|} \hline 
        Model & $\sigma(\log_{10}{(G_{\mathrm{eff}})})|_{\beta_{\phi}=1}$ & $\sigma(\log_{10}{(G_{\mathrm{eff}})})_{{\beta_{\phi}} \mathrm{free}}$  \\ \hline 
        SI$\nu$, $\log_{10}{(G_{\mathrm{eff}})} = 0$ & $\pm0.15 (>5)$ & $\pm0.95 (3.68)$ \\
        SI$\nu$, $\log_{10}{(G_{\mathrm{eff}})} = -1.5$ & $\pm0.37 (>5)$ & $\pm0.77 (2.59)$ \\
        MI$\nu$, $\log_{10}{(G_{\mathrm{eff}})} = -3.0$ & $\pm1.07 (0.46)$ & $\pm3.36 (0.0)$  \\  \hline 
        
    \end{tabular}
    \label{tab:CMB_forecasts_Planck}
\end{table}

The constraints here for $G_{\mathrm{eff}}$ in the CMB are comparable to those for neutrino self-interactions in other works \citep{kreisch2020neutrino, poudou2025self, camarena2025strong} without suffering from the degeneracy with $A_s$ and $n_s$ that requires the analysis to be split into two regions for SI and MI modes. Our analysis shows that with CMB data, the phase shift information may be a promising approach to measure or disfavour self-interactions that delay neutrino free-streaming. Similar to the case for the CMB, self-interactions will be poorly constraining in the case that $\log_{10}{(G_{\mathrm{eff}})} \leq -3$ due to the fact the derivative $\frac{dC(\ell)}{dG_{\mathrm{eff}}}$ approaches zero very quickly. However, it will likely be possible that model comparisons can help to disfavour strong interactions models with real data. 

\subsection{Combined BAO and CMB forecasts}\label{comboconstraints}

Finally, we consider the combined constraining power of BAO data from a galaxy survey with CMB data. This is mainly to test whether the phase shift information in the BAO is able to provide any useful additional constraining power for $G_{\mathrm{eff}}$ from the phase shift information available in a CMB experiment alone. In Table~\ref{tab:forecasts_PlanckDESI}, we show the forecasts for constraints on $G_{\mathrm{eff}}$ from the combined information in DESI YR5 BAO data with data from a \textit{Planck}-like CMB experiment. To compute the combined constraints, we take the Fisher information matrices from the forecasts for the BAO experiment from DESI, and use a Jacobian transformation to convert the Fisher matrix for the parameters $\{ \alpha_{i,\parallel}(z), \alpha_{i,\perp}(z), f\sigma_{i,8}(z), \beta_{\phi}, \log_{10}(G_{\mathrm{eff}}) \}$ (where $i$ is an index running over different redshift bins and tracers) to a new Fisher matrix with $\{100\theta_s, \Omega_{\mathrm{b}}h^2, \Omega_{\mathrm{c}}h^2, A_s, n_s, \tau, \beta_{\phi}, \log_{10}(G_{\mathrm{eff}}) \}$. Fisher information is additive, so in this form the Fisher information from the DESI data forecasts and the \textit{Planck}-like experiment can simply be added together. 

\begin{table}[h!]
    \caption{Fisher forecasts for the ability to constrain $G_{\mathrm{eff}}$, for a few different models, for a \textit{Planck}-like CMB measurement of the TT, EE and TE power spectra, combined with a measurement of data from DESI YR5. The columns are the same as those in Table~\ref{tab:BAOvolumelimitedresultsz0_2}; the second number show the relative improvement on the constraints from a CMB experiment alone. }
    \centering
    \begin{tabular}{|c|c|c|} \hline 
        Model & $\sigma(\log_{10}{(G_{\mathrm{eff}})})|_{\beta_{\phi}=1}$ & $\sigma(\log_{10}{(G_{\mathrm{eff}})})_{{\beta_{\phi}} \mathrm{free}}$  \\ \hline 
        SI$\nu$, $\log_{10}{(G_{\mathrm{eff}})} = 0$ & $\pm0.15$, 1.0 & $\pm0.93$,1.02 \\
        SI$\nu$, $\log_{10}{(G_{\mathrm{eff}})} = -1.5$ & $\pm0.35, 1.06$ & $\pm0.69, 1.11$ \\
        MI$\nu$, $\log_{10}{(G_{\mathrm{eff}})} = -3.0$ & $\pm 1.04, 1.03$ & $\pm 2.29, 1.47$  \\  \hline 
    \end{tabular}
    \label{tab:forecasts_PlanckDESI}
\end{table}

The results in Table~\ref{tab:forecasts_PlanckDESI} show that the BAO information is indeed able to improve the constraints by a factor of $\sim 1-1.47$ (up to a 47\% improvement) depending on the strength of the self-interactions. Since the CMB data is very constraining on its own for stronger self-interactions, the combined Fisher information from both probes together has the least improvement in this case. For weaker constraints from the CMB, the BAO information is more useful to help improve the constraints. The results here suggest that even though the BAO information alone may be less useful than CMB data for constraining self-interactions, BAO data can add useful information to the constraints from the phase shift that are worthwhile, especially for distinguishing weaker interactions.

\section{Conclusion}\label{sec:conclusion}

In summary, in this work we have explored the possibility of using the phase shift information from the BAO signal and CMB power spectra to constrain the strength of neutrino self-interactions. Due to such self-interactions in the early Universe, delayed neutrino free-streaming alters the typical phase shift signal expected from Standard Model neutrinos. 

We explore the phase shift, both in the BAO signal and the CMB power spectra, in the context of neutrinos with universal self-interactions parameterized by $G_{\mathrm{eff}}$. It is interesting that we show that the impact on the phase shift with increasing $G_{\mathrm{eff}}$ has a similar impact on the phase shift signal, both in the CMB and BAO, that is shown in the previous work of \cite{choi2018probing}. Ref.~\cite{choi2018probing} studies the CMB phase shift in the presence of free-streaming light relics with a delayed free-streaming epoch relative to Standard Model neutrinos, and shows that the CMB power spectra can constrain a delay in free-streaming light relics and consequently a change to $N_{\mathrm{eff}}$ in the context of a model that involves copies of the Standard Model (and thus additional neutrinos with delayed free-streaming). These results help to validate our own. 

We construct a template for the phase shift that captures the impact of altering $G_{\mathrm{eff}}$, both for the BAO oscillatory spectrum and the CMB power spectra, by building on the existing templates by \cite{baumann2018searching, baumann2019first} and \citep{montefalcone2025free}. Our new templates allow one to model the phase shift for both alterations to $N_{\mathrm{eff}}$ through $\beta_{\phi}$ and $\log_{10}{(G_{\mathrm{eff}}})$ through two parameters, $A(G_{\mathrm{eff}})$ and $B(G_{\mathrm{eff}})$. $A(G_{\mathrm{eff}})$ modulates the amplitude of the phase shift, while $B(G_{\mathrm{eff}})$ introduces a scale-dependent exponential damping into the phase shift signal. The templates can be used to simultaneously or separately fit $\beta_{\phi}$ and $\log_{10}{(G_{\mathrm{eff}}})$ on real data or to produce forecasts for the information available in only the phase shift, as we have done in this work. Our results show that both the BAO and the CMB are able to constrain $G_{\mathrm{eff}}$ from the phase shift information, but the constraining power is strongly dependent on $G_{\mathrm{eff}}$ itself. While the BAO constraints alone are quite weak in most cases and likely will not be useful for any realistic case in the future, the BAO does provide complementary information to CMB constraints, improving the constraining power by up to 47\% with the larger improvements corresponding to smaller $G_{\mathrm{eff}}$. On its own, a \textit{Planck}-like measurement of the phase shift in the CMB should be able to constrain $G_{\mathrm{eff}}$ with an uncertainty between $\pm 0.15 - 1.07$, for $\log_{10}{(G_{\mathrm{eff}})}$ between $[-3, 0.0]$. However, for weaker interactions, it is very difficult for BAO or CMB data to differentiate between weaker interaction strengths, since it is not possible to distinguish changes to the phase in such cases. As such, it will be likely that in this case, stronger interaction models can simply be disfavoured by the phase shift information in any data; only upper bounds on $\log_{10}{(G_{\mathrm{eff}})}$ may be possible. In future work, it will be possible to fit data from CMB experiments to constrain self-interactions from the phase shift as was done in ref.~\cite{montefalcone2025directly}, although their constraints are phrased in terms of a decoupling redshift rather than $G_{\mathrm{eff}}$. Another difference is that the scale dependence of the phase shift we include above which is neglected in ref.~\cite{montefalcone2025directly} given they find it to be a negligible effect in an analysis on current CMB data. They also considers a broader range of temperature dependence for the neutrino interactions than we do, as implemented in their \texttt{nuCLASS}\footnote{\url{https://github.com/subhajitghosh-phy/nuCLASS.git}} Boltzmann solver. In their Appendix B, one can see how changing the temperature dependence of the interactions impacts the phase shift signature, with the scale dependence being negligible for weaker temperature scaling but becoming more significant for interactions with a steeper temperature dependence.  The temperature dependence for the interaction rate is also studied in \cite{pal2025exploring}. It would be interesting to determine
how the subtle scale dependence we have included in our phase-shift template impacts the constraints on the neutrino decoupling redshift and $G_{\mathrm{eff}}$.   

Overall, it is interesting to study how different neutrino interactions may change the phase in the CMB or BAO, which is a robust signature of the C$\nu$B, and constraints from cosmological data can be complementary to particle physics experiments which may not be able to access interactions between particles in the conditions that occur in the early Universe. As opposed to studying the general effects of neutrino self-interactions via the $G_{\mathrm{eff}}$ parameterization which changes the broadband shape of the matter power spectrum in a manner that may lead to strong degeneracies with $A_s$ and $n_s$, studying the phase shift alone may help improve constraints by isolating a robust impact of neutrino species that free-stream at early times or instead decouple at a delayed time in the case of self-interactions. While we found obtaining strong constraints from the BAO alone would be extremely challenging, there is potential for interesting constraints from CMB data (and combinations of this data with BAO information). 

There may also be additional information from the phase shift that could be explored in other probes. In future work, it may be worth exploring forecasts in which the helium abundance $Y_{p}$, which describes the mass abundance of helium, could be included in the CMB forecasts. This may weaken constraints because $Y_{p}$ changes the Silk damping scale and thus is expected to have some degeneracy with $\beta_{\phi}(N_{\mathrm{eff}})$. In \cite{libanore2025joint}, it is shown that the 21 cm power spectrum is able to constrain $G_{\mathrm{eff}}$; these constraints could be complementary to those obtained from the phase shift in CMB and BAO data. In \cite{montefalcone2025tracing}, the phase shift is studied in the 21 cm power spectrum.

Various interesting extensions could be considered for this work. Firstly, this work only explores the phase shift as it arises for neutrinos with universal self-interactions; a model that is largely disfavored by experimental data \citep{lyu2021self, blinov2019constraining}. However, it is possible to have non-universal self-interactions, and it might be expected this would change the significance of the effect self-interactions has on the phase shift if only some neutrinos have delayed free-streaming. In future work, it may be possible to develop a template for such models. Interestingly, \cite{das2021flavor} shows fits to \textit{Planck} data for $G_{\mathrm{eff}}$ in which only one or two neutrino species self-interact, and this provides a better fit to the data than the case of three neutrinos self-interacting; in contrast, \cite{brinckmann2021self} does not find any preference for neutrino self-interactions when studying varying numbers of self-interacting or free-streaming species. Another possible extension could be to consider models of resonant neutrino self-interactions \citep{noriega2025resonant} or models in which neutrinos `self-recouple' at a later time instead of having delayed free-streaming \citep{berryman2023neutrino}. In the case of resonant self-interactions explored in \citep{noriega2025resonant}, for specific mediator masses $m_{\phi}$, the neutrinos self-interact preferably at $z \sim 4 \times 10^4 \frac{m_{\phi}}{eV}$, which corresponds to the epoch of BAO propagation. Consequently, the signal of self-interactions can be boosted, and we might expect this to also have a different impact on the phase shift that could be explored. Additionally, for some mediator masses it may be possible for neutrinos to `self-recouple'; in this case, modes that enter the causal horizon before the neutrinos self-recouple will experience the same phase shift that we expect to see for Standard Model neutrinos, while those that enter afterwards may have a similar impact on the phase shift that we see in the case for self-interactions explored in this work.

\newpage 

\appendix

\newpage 

\section{The effect of $N_{\mathrm{eff}}$ on the phase shift}\label{App:B}

\begin{figure}[h!]
\centering
    \begin{subfigure}{0.8\textwidth} \includegraphics[width=\linewidth]{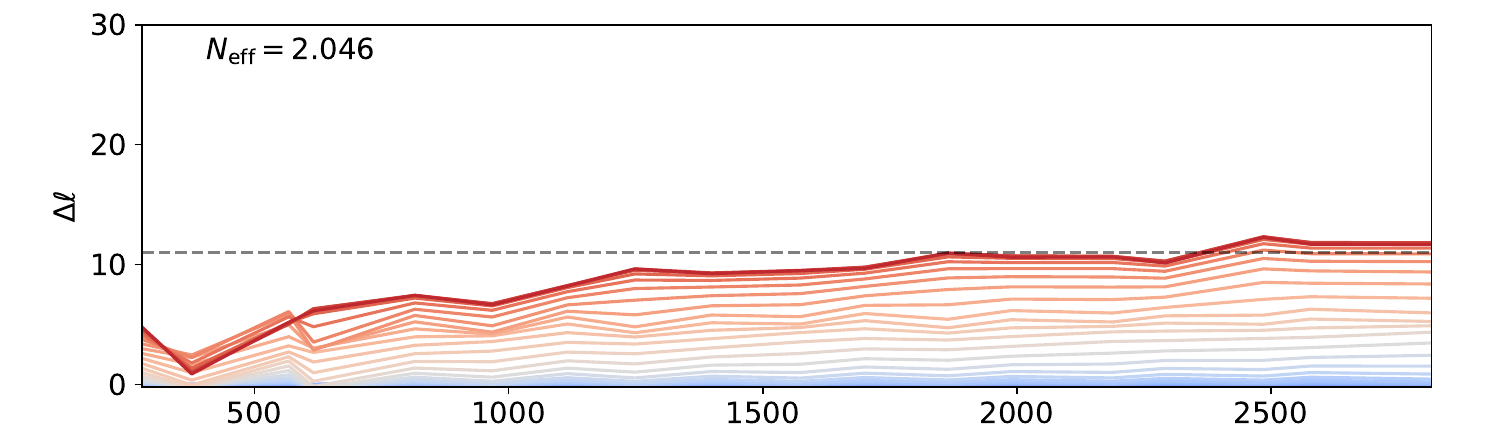}
    \end{subfigure}
    \begin{subfigure}{0.8\textwidth} \includegraphics[width=\linewidth]{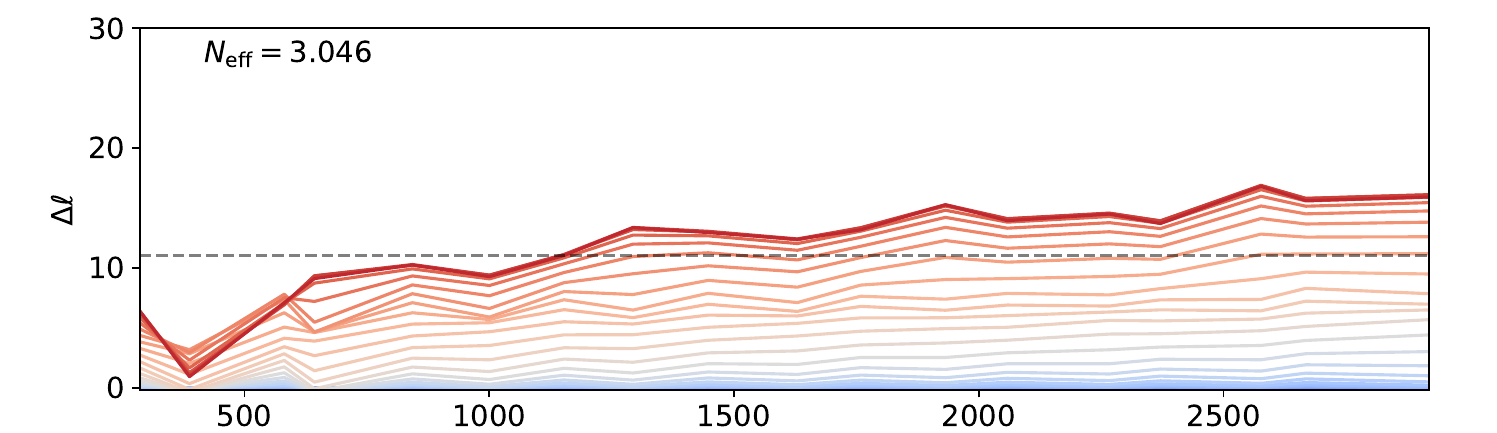}
    \end{subfigure}
    \begin{subfigure}{0.8\textwidth} \includegraphics[width=\linewidth]{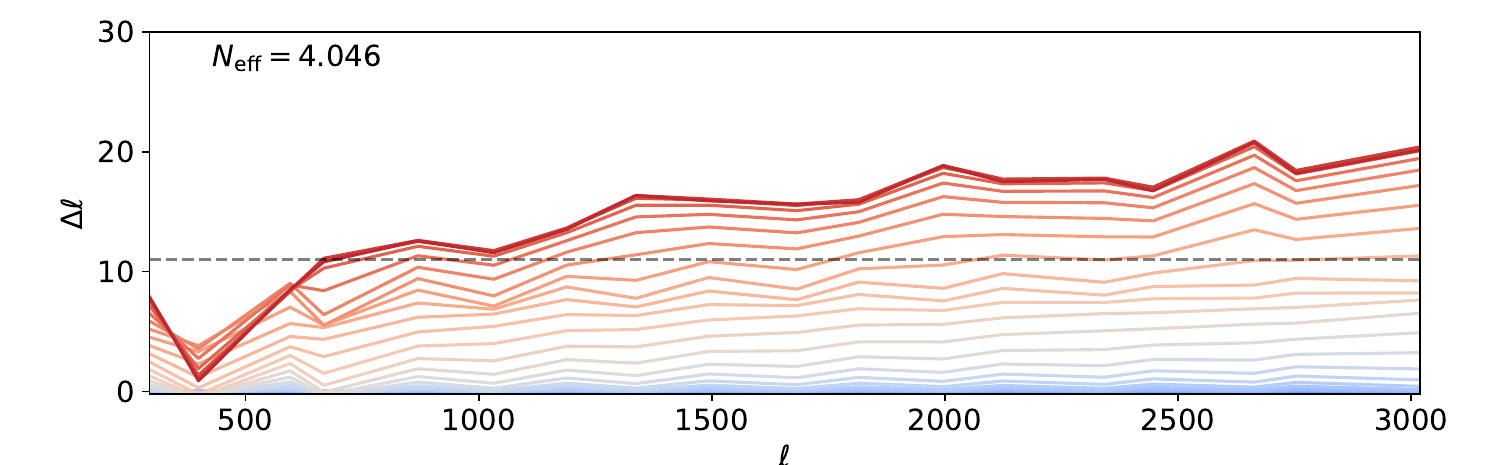}
    \end{subfigure}
    \begin{subfigure}{0.8\textwidth} \includegraphics[width=\linewidth]{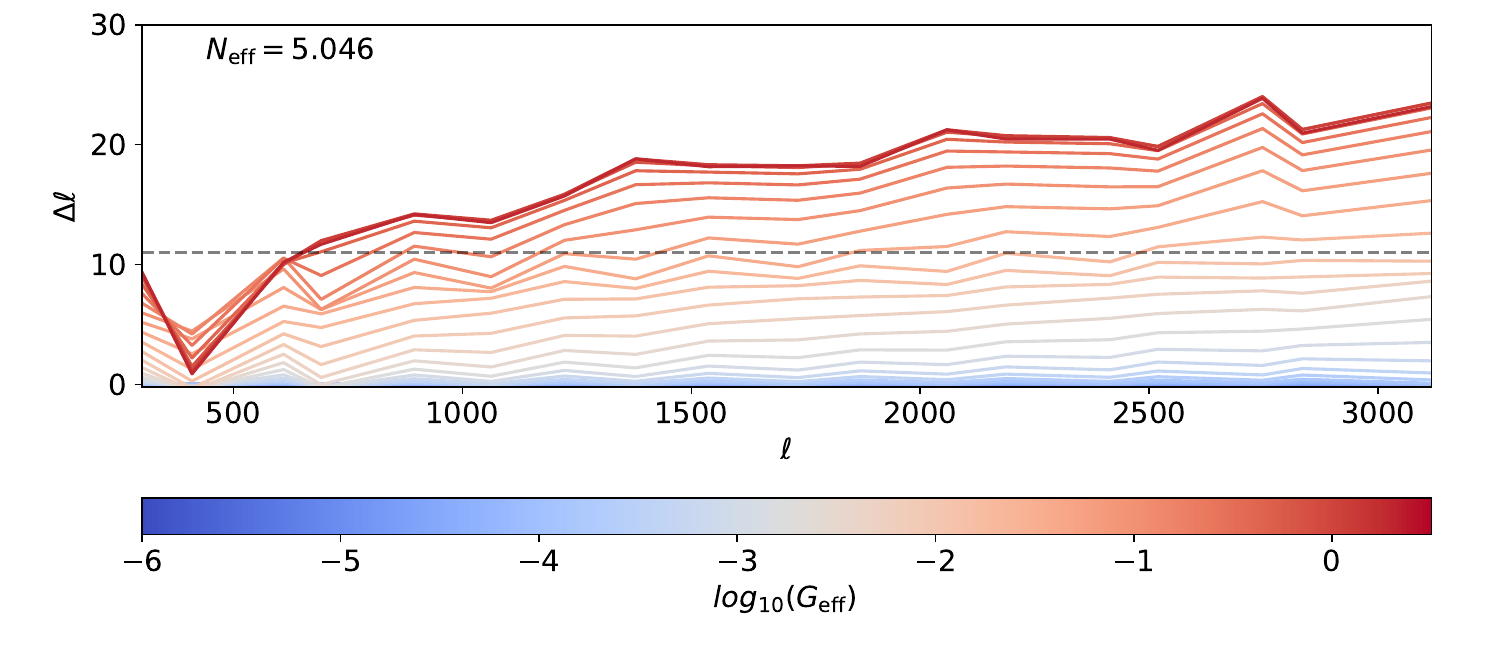}
    \end{subfigure}
    \caption{The phase shift $\Delta \ell$, demonstrated in the CMB temperature-temperature power spectrum, when varying $\log_{10}{(G_{\mathrm{eff}})}$. The various panels show the phase shift for different fixed values of $N_{\mathrm{eff}}$, to demonstrate the impact that additional light relics have on the amplitude of the phase shift, and to show that it is possible to separate the dependence of the phase shift signal on $N_{\mathrm{eff}}$ and $\log_{10}{(G_{\mathrm{eff}})}$. }\label{fig:neffeffectphase}
\end{figure}

The effect of $N_{\mathrm{eff}}$ on the phase shift amplitude is demonstrated in Figure~\ref{fig:neffeffectphase}; for neutrinos with universal self-interactions, a larger value of $N_{\mathrm{eff}}$ increases the impact of the phase shift induced by self-interactions, relative to a model with no self-interactions. The goal here is to show that it is possible to separate the impacts of $N_{\mathrm{eff}}$ and $G_{\mathrm{eff}}$ on the phase shift. However, we note that in general for a fixed value of $G_{\mathrm{eff}}$, the epoch at which the neutrinos decouple may slightly vary when $N_{\mathrm{eff}}$ is varied. Therefore, the approach in \cite{montefalcone2025directly} was to study the phase shift as a function of the decoupling redshift of the neutrinos rather than $G_{\mathrm{eff}}$, although both approaches are valid. 

\begin{figure}
    \centering
    \includegraphics[width=1\linewidth]{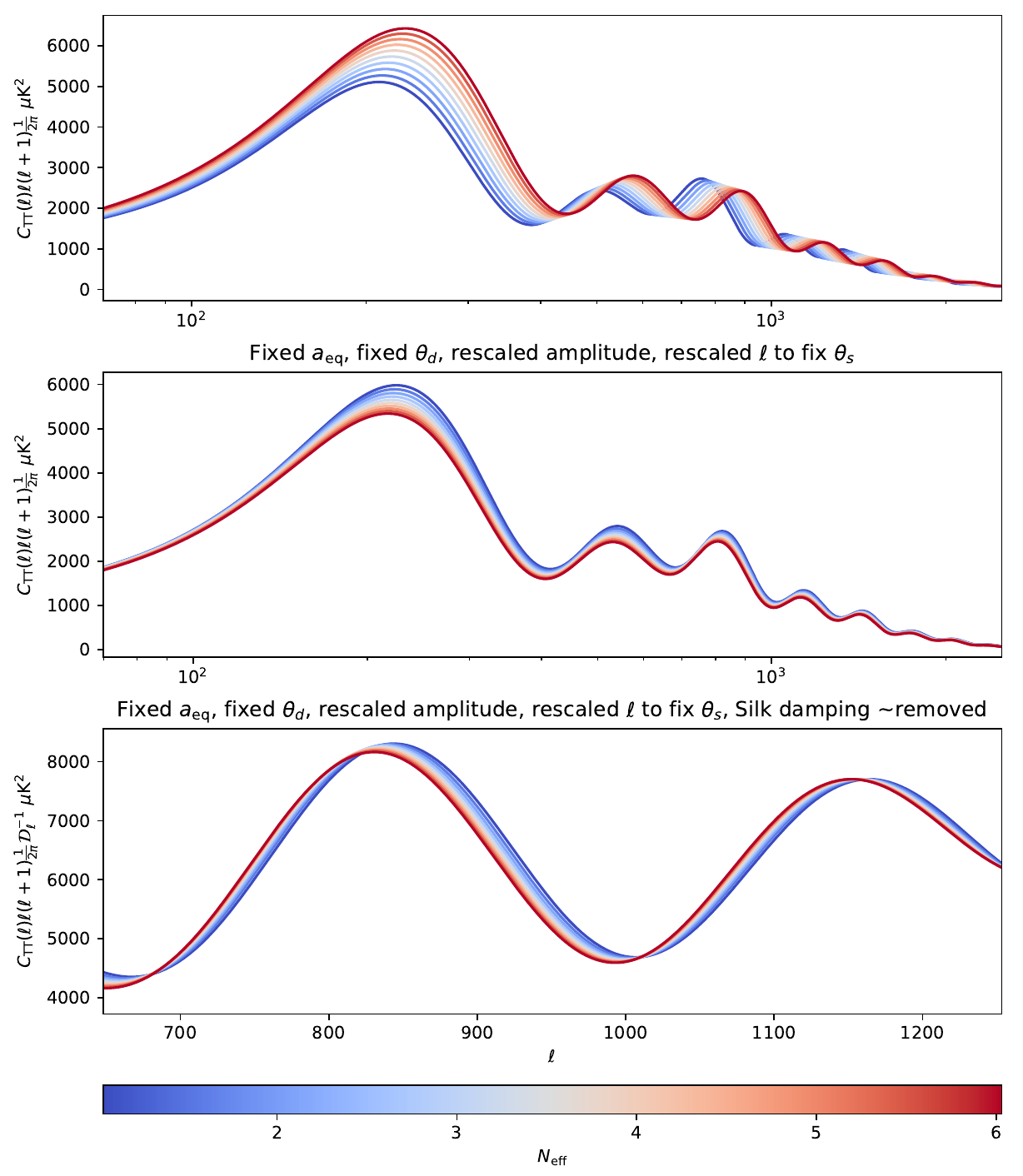}\caption{The neutrino-induced phase shift in the CMB. The top panel is the CMB with changing $N_{\mathrm{eff}}$. In the middle panel, the phase shift has been isolated by removing the impact of changes to $a_{\mathrm{eq}}$, $\theta_s$, $\theta_D$ and changes to the amplitude of the CMB. The lowest panel shows the same but has been undamped (approximately) by multiplying by $\sim e^{0.25 (\ell \theta_D)^{1.35}}$ following the approach taken in \cite{follin2015first, montefalcone2025free} to approximately remove the impact of Silk damping; this exponential damping is cosmology dependent. This shows the isolated phase shift for standard model neutrinos. }\label{fig:CMB_standardneffphaseshift}
\end{figure}

In Figure~\ref{fig:CMB_standardneffphaseshift} we show the impact of $N_{\mathrm{eff}}$ on the CMB temperature power spectrum which has a measurable phase shift (both in the CMB and BAO signal) independent of $G_{\mathrm{eff}}$, due to neutrino free-streaming. This has been previously studied \cite{bashinsky2004neutrino, green2020phase, baumann2018searching} and measured in \cite{follin2015first, baumann2019first, whitford2024constraining, montefalcone2025free}. 

\acknowledgments

This research has made use of NASA's Astrophysics Data System bibliographic services, the astro-ph pre-print archive at \url{https://arxiv.org/} and the python libraries \textsc{MATPLOTLIB, NUMPY, SCIPY} and \textsc{PANDAS} \citep{harris2020array, mckinney2010proceedings, virtanen2020scipy, 2020SciPy-NMeth, mckinney-proc-scipy-2010, hunter2007matplotlib}. AW thanks the Galileo Galilei Institute of Theoretical Physics in Florence, Italy, for facilitating discussion that led to the research conducted in this paper at the \textit{Neutrino Frontiers} workshop in 2024. AW also thanks the University of Queensland and Australian Government for funding through the Australian Government Research Training Program Stipend. AW, CH and TD acknowledge support from the Discovery Project (project DP20220101395) funding scheme and Australian Research Council Centre of Excellence for Gravitational Wave Discovery (OzGrav), through project number CE230100016.  D. C.~and F.-Y. C.-R.~would like to thank the Robert E.~Young Origins of the Universe Chair fund for its generous support. F.-Y. C.-R. acknowledges the support of the US National Science Foundation CAREER grant AST-2440096. We also thank Gabriele Montefalcone and the anonymous reviewer for their useful comments which helped to improve this manuscript. 

% \paragraph{Note added.} This is also a good position for notes added
% after the paper has been written.


\bibliographystyle{JHEP.bst}
\bibliography{bibtemplate}

% \begin{thebibliography}{99}

% \bibitem{a}
% Author, \emph{Title}, \emph{J. Abbrev.} {\bf vol} (year) pg.

% \bibitem{b}
% Author, \emph{Title},
% arxiv:1234.5678.

% \bibitem{c}
% Author, \emph{Title},
% Publisher (year).

% Please avoid comments such as "For a review'', "For some examples",
% "and references therein" or move them in the text. In general,
% please leave only references in the bibliography and move all
% accessory text in footnotes.

% Also, please have only one work for each \bibitem.

% \end{thebibliography}

\end{document}